\documentclass[conference, letterpaper]{IEEEtran}
\usepackage{comment}
\usepackage{placeins}
\usepackage{algorithm,algorithmic}
\usepackage{amsmath,amssymb,latexsym,amsmath,euscript,mathrsfs}
\usepackage[pdftex]{graphicx}
\usepackage{subfigure}
\usepackage{multirow}
\usepackage{color}
\usepackage{url}
\usepackage[compress]{cite}

\begin{document}

\newcommand{\fig}[1]{Fig.~\ref{#1}}
\newcommand{\tab}[1]{Table~\ref{#1}}
\newcommand{\sect}[1]{Section~\ref{#1}}
\newcommand{\appx}[1]{Appendix~\ref{#1}}
\newcommand{\eqn}[1]{Eq.~\ref{#1}}
\newcommand{\alg}[1]{Algorithm~\ref{#1}}

\newcommand{\eg}{\textit{e.g.},} 
\newcommand{\ie}{\textit{i.e.}, }
\newcommand{\etc}{\textit{etc}.}
\newcommand{\wrt}{\textit{w.r.t.} }
\newcommand{\etal}{\textit{et al.} }

\title{On Orchestrating Virtual Network Functions in NFV}

\author{
	\IEEEauthorblockN{
		Md. Faizul Bari,
		Shihabur Rahman Chowdhury,
		Reaz Ahmed, and
		Raouf Boutaba
	}

	\IEEEauthorblockA{
		David R. Cheriton School of Computer Science, University of Waterloo \\
		\texttt{[mfbari | sr2chowdhury | r5ahmed | rboutaba]@uwaterloo.ca}
	}
}

\maketitle

\begin{abstract}
Middleboxes or network appliances like firewalls, proxies and WAN optimizers have become an integral part of today's ISP and enterprise networks. Middlebox functionalities are usually deployed on expensive and proprietary hardware that require trained personnel for deployment and maintenance. Middleboxes contribute significantly to a network's capital and operational costs. In addition, organizations often require their traffic to pass through a specific sequence of middleboxes for compliance with security and performance policies. This makes the middlebox deployment and maintenance tasks even more complicated. Network Function Virtualization (NFV) is an emerging and promising technology that is envisioned to overcome these challenges. It proposes to move packet processing from dedicated hardware middleboxes to software running on commodity servers. In NFV terminology, software middleboxes are referred to as Virtualized Network Functions (VNFs). It is a challenging problem to determine the required number and placement of VNFs that optimizes network operational costs and utilization, without violating service level agreements. We call this the VNF Orchestration Problem (VNF-OP) and provide an Integer Linear Programming (ILP) formulation with implementation in CPLEX. We also provide a dynamic programming based heuristic to solve larger instances of VNF-OP. Trace driven simulations on real-world network topologies demonstrate that the heuristic can provide solutions that are within 1.3 times of the optimal solution. Our experiments suggest that a VNF based approach can provide more than $\mathbf{4\times}$ reduction in the operational cost of a network.
\end{abstract}

\section{Introduction}\label{sec:introduction}
Today's enterprise networks ubiquitously deploy vertically integrated proprietary middleboxes or network appliances to offer various network services. Examples of such middleboxes include firewalls, proxies, WAN optimizers, Intrusion Detection Systems (IDSs) and Intrusion Prevention Systems (IPSs). A recent study shows that the number of different middleboxes is comparable to the number of routers in an enterprise network~\cite{Sherry:EECS-2012-24}. However, middleboxes come with high Capital Expenditures (CAPEX) and Operational Expenditures (OPEX). They are usually expensive, vendor specific and require specially trained personnel for deployment and maintenance. Moreover, it is often impossible to add new functionality to an existing middlebox, which makes it very difficult for the network operators to deploy new services. In many cases, network operators are compelled to purchase new hardware that substantially increases their CAPEX. 

Another set of problems arise from the fact that most often a traffic is required to pass through multiple stages of middlebox processing in a particular order, \eg\ a traffic may be required to go through an IDS, then a proxy and finally through a firewall~\cite{qazi2013simple}. This phenomenon is very common for middleboxes and is typically referred to as Service Function Chaining (SFC)~\cite{quinn2014service}. The IETF Network and Service Chaining Working Group has several IETF drafts demonstrating middlebox chaining use-cases in operator networks~\cite{liu2014service}, mobile networks~\cite{haeffner2014mobile} and data center networks~\cite{surendra2014dc}. The task of sequencing these in-network processing is commonly referred to as \emph{middlebox orchestration}. Currently, this task is performed by crafting the routing table entries manually. It is a cumbersome and error-prone process. Moreover, any placement of these hardware middleboxes is bound to become inefficient over time. It is very expensive and inconvenient to keep changing the locations of these hardware middleboxes with changing network conditions.

An emerging and promising technology that can address these limitations is Network Function Virtualization (NFV)~\cite{etsinfv}. It proposes to move packet processing from hardware middleboxes to software. Instead of running hardware based middleboxes, the same packet processing tasks are performed by software middleboxes running on commodity (\eg\ \texttt{x86} based systems) servers. Researchers have already developed virtualized platforms for software based middlebox processing that can achieve near hardware performance~\cite{martins2014clickos,179739}. In NFV terminology, these software middleboxes are referred to as Virtualized Network Functions (VNFs). NFV is envisioned to solve most of the above mentioned problems with hardware middleboxes. NFV also provides opportunities for network optimization and cost reduction. Previously, middleboxes were hardware appliances placed at fixed locations, but we can deploy a VNF virtually anywhere in the network. This feature opens-up a whole new window of opportunity to reduce both CAPEX and OPEX. Network operators no longer need to buy specialized hardware. This would significantly reduce CAPEX. In addition, VNFs can be orchestrated autonomically without requiring specially trained personnel for deployment and maintenance, which would reduce the operational and maintenance costs. 

VNFs can be orchestrated by deploying a composition of VNFs either on the same server or on a cluster of servers. The locations of the VNFs must be chosen carefully to ensure that traffic can be routed through them in the proper sequence with minimal changes in the forwarding tables. An emerging technology that can assist in flexible routing is Software Defined Networking (SDN)~\cite{kreutz2014software}. SDN decouples the control plane from the data plane and places it on a logically centralized controller. SDN control plane has a global network view and can be used to programmatically configure forwarding rules in the switches/routers to enable VNF orchestration.

VNFs promise to reduce the CAPEX and OPEX of a network. However, several issues need to be considered before provisioning VNFs: (i) the cost of deploying a new VNF, (ii) energy cost for running a VNF and (iii) the cost of forwarding traffic to and from a VNF. Placing just enough VNFs to match processing requirements of traffic may yield the lowest deployment and energy cost, but steering traffic through these VNFs will result in the use of sub-optimal paths that may lead to Service Level Objective (SLO) violations and customer dissatisfaction. On the other hand, one may try to always forward traffic through the shortest possible path by deploying VNFs whenever needed. This approach may avoid SLO violation penalty, but will surely lead to huge deployment and energy cost. An optimal VNF placement strategy is needed to find a suitable point between these two extreme cases, \ie a strategy that minimizes OPEX, penalty for SLO violations and resource fragmentation as well as forwards traffic through the best available path. We refer to this problem as the \textit{Virtualized Network Function Orchestration Problem (VNF-OP)}. 

Our key contributions can be summarized as follows: 
\begin{itemize}
\item We identify the VNF orchestration problem and provide the first quantifiable results showing that dynamic VNF orchestration can have more than $4\times$ reduction in OPEX. 
\item The problem is formulated as an Integer Liner Program (ILP) and implemented in CPLEX to find optimal solutions for small scale networks. 
\item Finally, we propose a fast heuristic algorithm that can find solutions within 1.3 times of the optimal. Its performance is evaluated using real-world topologies and traffic traces.
\end{itemize}

The rest of the paper is organized as follows: we start by explaining the mathematical model used for our system and by formally defining the VNF Orchestration Problem (\sect{sec:model}). Then the problem formulation is presented (\sect{sec:formulation}). Next, a heuristic is proposed to obtain near-optimal solutions (\sect{sec:heuristic}). We validate our solution through trace driven simulations on real-world network topologies (\sect{sec:evaluation}). Then, we provide a literature review (\sect{sec:related_work}) and finally, we conclude with some future directions (\sect{sec:conclusion}).
\section{Mathematical Model and Problem Definition}\label{sec:model}
In this section we introduce the mathematical model for our system and formally define the VNF Orchestration Problem. 

\subsection{Physical Network}
We represent the physical network as an undirected graph $\bar{G}=(\bar{S}, \bar{L})$, where $\bar{S}$ and $\bar{L}$ denote the set of switches and links, respectively. We assume that VNFs can be deployed on commodity servers located within the network. The set $\bar{N}$ represents these servers and the binary variable $\bar{h}_{\bar{n}\bar{s}} \in \{0, 1\}$ indicates whether server $\bar{n} \in \bar{N}$ is attached to switch $\bar{s} \in \bar{S}$.

\begin{align*}
\bar{h}_{\bar{n}\bar{s}} & =\left\{ \begin{array}{rl}
1 & \text{if server } \bar{n} \in \bar{N} \text{ is attached to switch } \bar{s} \in \bar{S},\\
0 & \text{otherwise}.
\end{array}\right.
\end{align*}

Let, $R$ denote the set of resources (CPU, memory, disk, \etc) offered by each server. The resource capacity of server $\bar{n}$ is denoted by $c^{r}_{\bar{n}} \in \mathbb{R}^{+},\ \forall\ r \in R$. The bandwidth capacity and propagation delay of a physical link $(\bar{u}, \bar{v}) \in \bar{L}$ is represented by $\beta_{\bar{u}\bar{v}} \in \mathbb{R}^{+}$ and $\delta_{\bar{u}\bar{v}} \in \mathbb{R}^{+}$, respectively. We also define a function $\eta(\bar{u})$ that returns the neighbors of switch $\bar{u}$.

\begin{align*}
\eta(\bar{u}) = \{\bar{v}\ |\ (\bar{u}, \bar{v}) \in \bar{L} \text{ or } (\bar{v}, \bar{u}) \in \bar{L} \},\ \bar{u}, \bar{v} \in \bar{S}
\end{align*}

\subsection{Virtualized Network Functions (VNFs)}
Different types of VNFs (\eg\ firewall, IDS, IPS, proxy, \etc) can be provisioned in a network. Set $P$ represents the possible VNF types. VNF type $p$ has a specific deployment cost, resource requirement, processing capacity and processing delay represented by $\mathcal{D}^{+}_{p}$, $\kappa_{p}^r \in \mathbb{R}^+ (\forall r \in R)$, $c_p$ (in Mbps) and $\delta_p$ (in ms), respectively. 

There can be certain hardware requirements (\eg\ hardware-accelerated encryption for Deep Packet Inspection (DPI)) that may prevent a server from running a particular type of VNF. Furthermore, the network manager may have preferences regarding provisioning a particular type of VNF on a particular set of servers, \eg\ Firewalls should be run close to the edge of the network. So, we assume that for each VNF type there is a set of servers on which it can be provisioned. The following binary variable represents this relationship:

\begin{align*}
d_{\bar{n}p} & =\left\{ \begin{array}{rl}
1 & \text{if VNF type } p \in P \text{ can be provisioned on } \bar{n},\\
0 & \text{otherwise}.
\end{array}\right.
\end{align*}

\subsection{Traffic Request}
We assume that the network operator is receiving requests for setting up paths for different kinds of traffic (\eg\ VPN setup, expected traffic for a new application or service in a data center, \etc). A traffic request is represented by a 6-tuple $t = \langle \bar{u}^t, \bar{v}^t, \Psi^t, \beta^t, \delta^t, \omega^t \rangle$, where $\bar{u}^t, \bar{v}^t \in \bar{S}$ denote the ingress and egress switches, respectively. $\beta^t \in \mathbb{R}^+$ is the bandwidth demand of the traffic. $\delta^t$ is the expected propagation delay according to Service Level Agreement (SLA). $\Psi^t$ represents the ordered VNF sequence the traffic must pass through (\eg\ Firewall $\rightarrowtail$ IDS $\rightarrowtail$ Proxy). $l_{\Psi^t}$ denotes the length of $\Psi^t$ and $\omega^t$ denotes policy to determine SLO violation penalties. 

\begin{figure}[!htbp]%
	\centering%
		\includegraphics[width=0.4\textwidth]{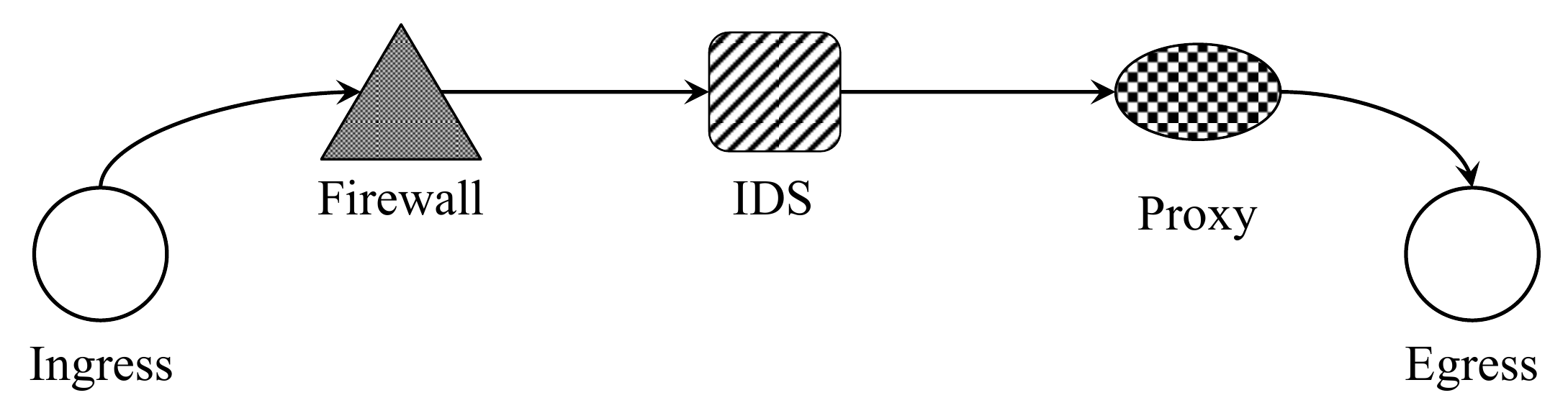}%
		\caption{Traffic Model}%
	\label{fig:traffic_model}%
\end{figure}%

We represent a traffic request $t$ by a directed graph $G^t = (N^t, L^t)$, where $N^t$ represents the set of \emph{traffic nodes} (switches and VNFs) and $L^t$ denotes the links between them.~\fig{fig:traffic_model} represents a traffic that requires to pass through the VNF sequence: Firewall $\rightarrowtail$ IDS $\rightarrowtail$ Proxy. Modeling the traffic in this way makes it easy for the provisioning process to ensure that it passes though the correct sequence of VNFs. We also define  $\eta^t(n_1)$ to represent the neighbors of $n_1 \in N^t$:

\begin{align*}
\eta^t(n_1) = \{n_2\ |\ (n_1, n_2) \in L^t\},\ n_1, n_2 \in N^t
\end{align*}

Next, we define a binary variable $g^t_{np} \in \{0, 1\}$ to indicate the type of a node $n \in N^t$

\begin{align*}
g^t_{np} & =\left\{ \begin{array}{rl}
1 & \text{if node } n \in N^t \text{ is of type } p \in P,\\
0 & \text{otherwise}.
\end{array}\right.
\end{align*}

\subsection{VNF Orchestration Problem (VNF-OP)}
We consider a scenario where an operational network is serving a set of traffics $\hat{T}$. It has a set of VNFs already deployed and the routing paths for the traffics in $\hat{T}$ are also provisioned. Now, the network operator is receiving new traffic requests and wants to provision the required VNFs and routing paths for them. The network operator can choose to provision resources for one traffic request at a time or leverage a lookahead interval by accumulating a number of traffic requests and provision resources in batches. Determining the optimal number or volume of traffic or the length of the lookahead interval for each batch is an interesting research challenge that we consider out-of-scope for the current work and plan to pursue in the future. In the rest of the paper, we denote a new traffic batch by $T$. Based on the operator's choice, a batch may contain just one or multiple traffic requests. 

In the VNF-OP, we are given a physical network topology, VNF specifications, current network status and a set of new traffic requests. Our objective is to minimize the overall network OPEX and physical resource fragmentation by (i) provisioning an optimal number of VNFs, (ii) placing them at the optimal locations and (ii) finding the optimal routing paths for each traffic request, while respecting the capacity constraints (\eg\ physical servers, links, and VNFs) and ensuring that traffic passes through the proper VNF sequence.

\textbf{-- OPEX:}
In this work, we consider the network OPEX to be composed of the following four cost components:
\begin{itemize}
\item \text{\emph{VNF deployment cost:} } we need to complete tasks like transferring a VM image, booting it and attaching it to devices before deploying a VNF. We associate a cost (in dollars) with these operations. 
\item \text{Energy cost: } it represents the cost of energy consumption for the active servers. A server is considered active if it has at least one active VNF. Servers consume power based on the amount of resources (\eg\ CPU, memory, disk, \etc) under use. A server is assumed to be in the idle state if it does not have any active VNFs~\cite{serverpower}.
\item \text{Traffic forwarding cost: } an operator incurs traffic forwarding cost from two sources: (i) leasing cost of transit links~\cite{transitcost} and (ii) energy consumption of the network devices (\eg\ switches, routers, \etc). 
\item \text{Penalty for SLO violation: } this cost component represents the penalty that must be paid to the customer for SLO violations, \eg\ if a traffic experienced more that the maximum allowed propagation delay.
\end{itemize}

\textbf{-- Resource Fragmentation:}
We compute the physical resource fragmentation by measuring the percentage of idle resources for the active servers and links. We want to minimize fragmentation as it eventually increases the possibility of accommodating more traffic on the same resource.
\section{Integer Linear Programming (ILP) Formulation}\label{sec:formulation}
VNF-OP is a considerably harder problem to solve than traditional Virtual Network (VN) embedding problems~\cite{Mosharaf10}. There is no node ordering requirement in VN embedding, while in VNF-OP we need to preserve the ordering of VNFs. Moreover, in VNF-OP we need to respect the processing capacity constraints of servers and the VNFs to be deployed. How many VNFs are to be deployed is not known in advance, rather it is an outcome of the optimization process. Multi-dimensional Bin Packing~\cite{kou1977multidimensional} can also be used to solve VNF-OP, but here we will end-up with a \emph{nested} bin packing problem. In the first layer traffics need to be packed into VNFs and in the next layer VNFs need to be packed into the physical servers. The fact that the number and locations of VNFs is not known in advance, results in quadratic constraints for resource capacity and renders the problem unsolvable even for very small instances by existing optimization solvers. In this work, we address these challenges by judiciously augmenting the physical network, which is explained in the rest of the section. 

\subsection{Physical Network Transformation}\label{subsec:trandformation}
We transform the physical network to generate an augmented pseudo-network that reduces the complexity involved in solving the VNF-OP. The transformation process is performed in two steps:

\begin{figure*}
	\centering
	\subfigure[Original Network]{\includegraphics[width=0.27\textwidth]{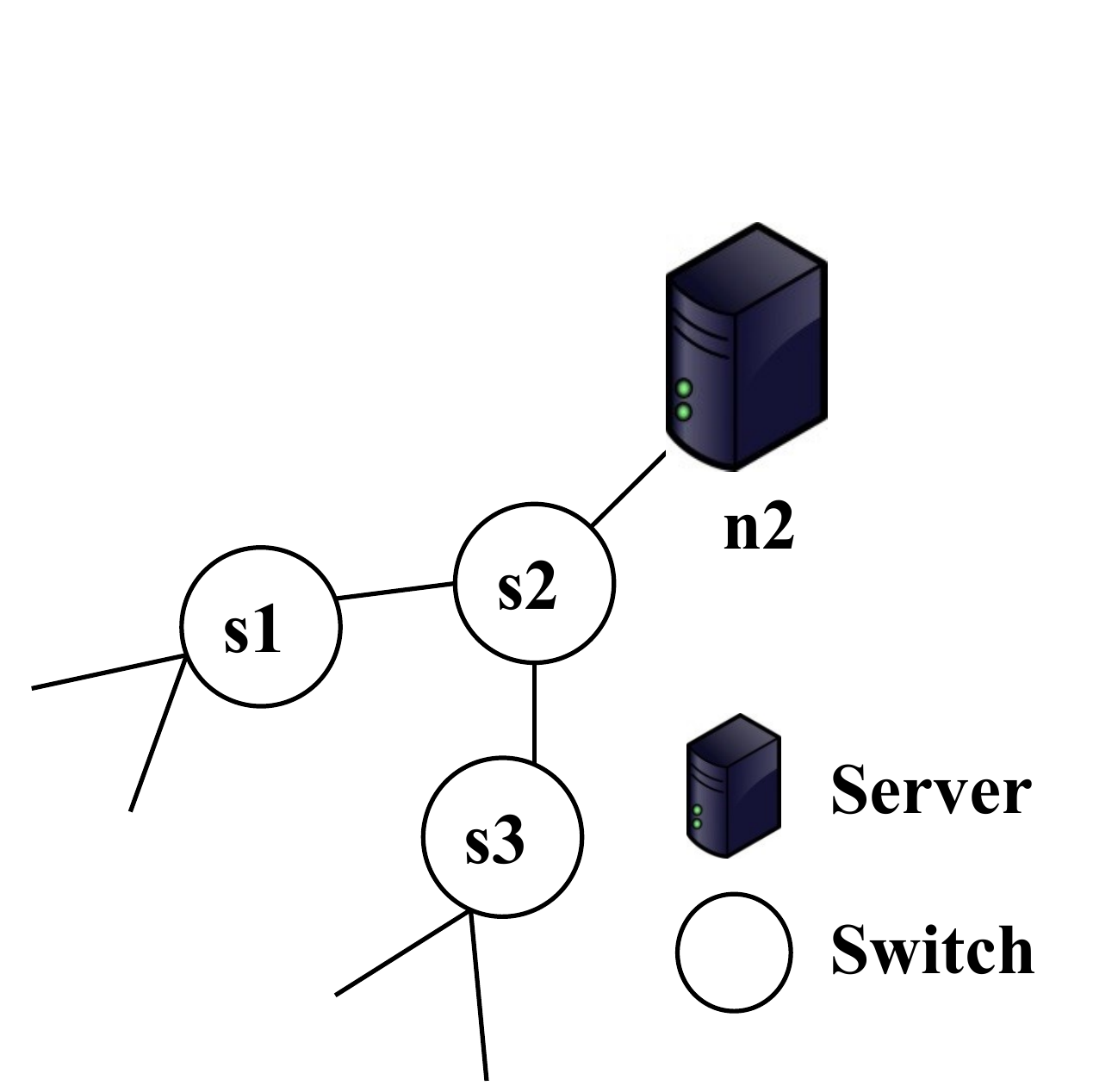}\label{fig:initial_network}}
	\subfigure[VNF Enumeration]{\includegraphics[width=0.27\textwidth]{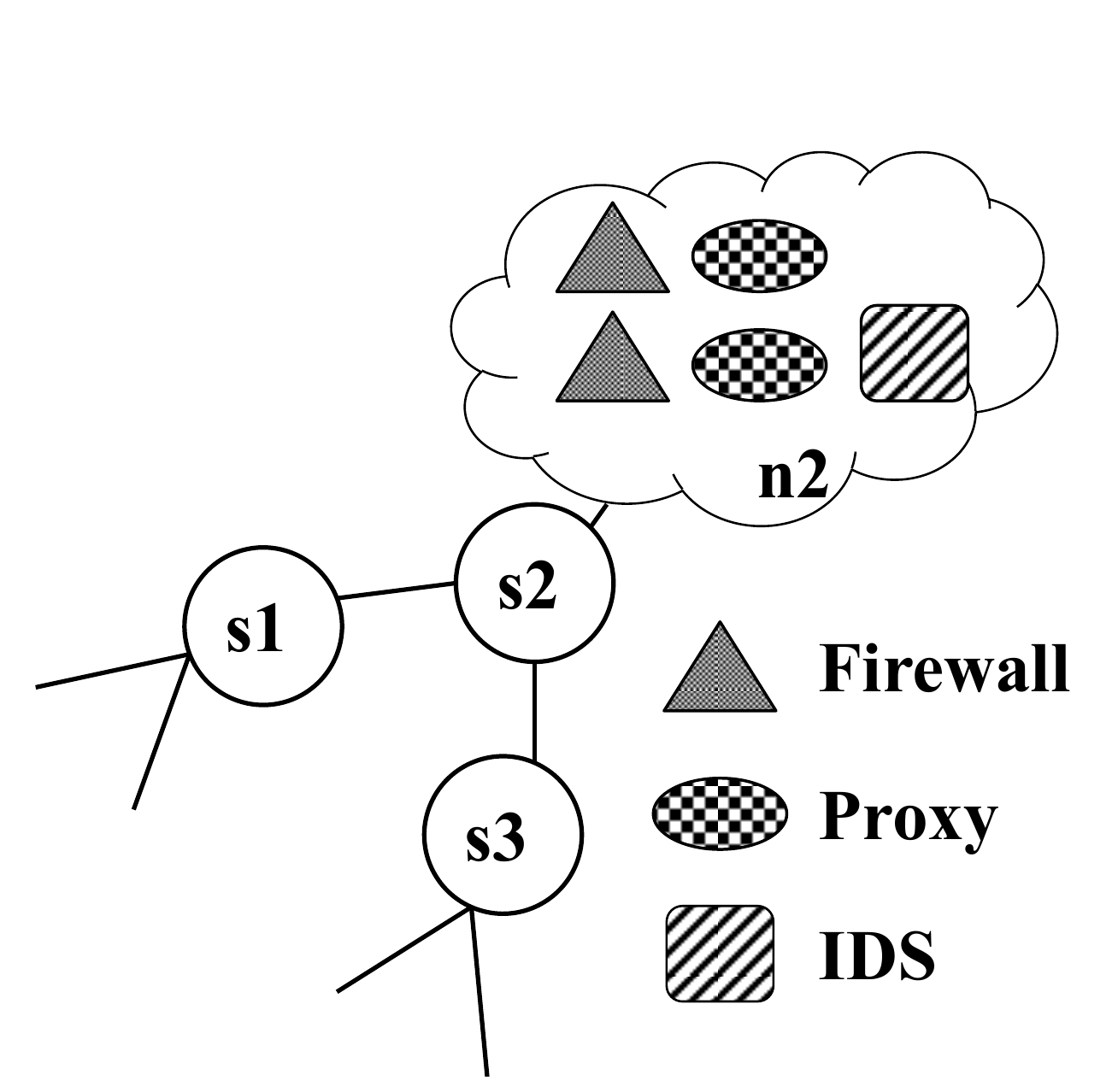}\label{fig:vnf_enumeration}}
	\subfigure[Adding pseudo-switches]{\includegraphics[width=0.27\textwidth]{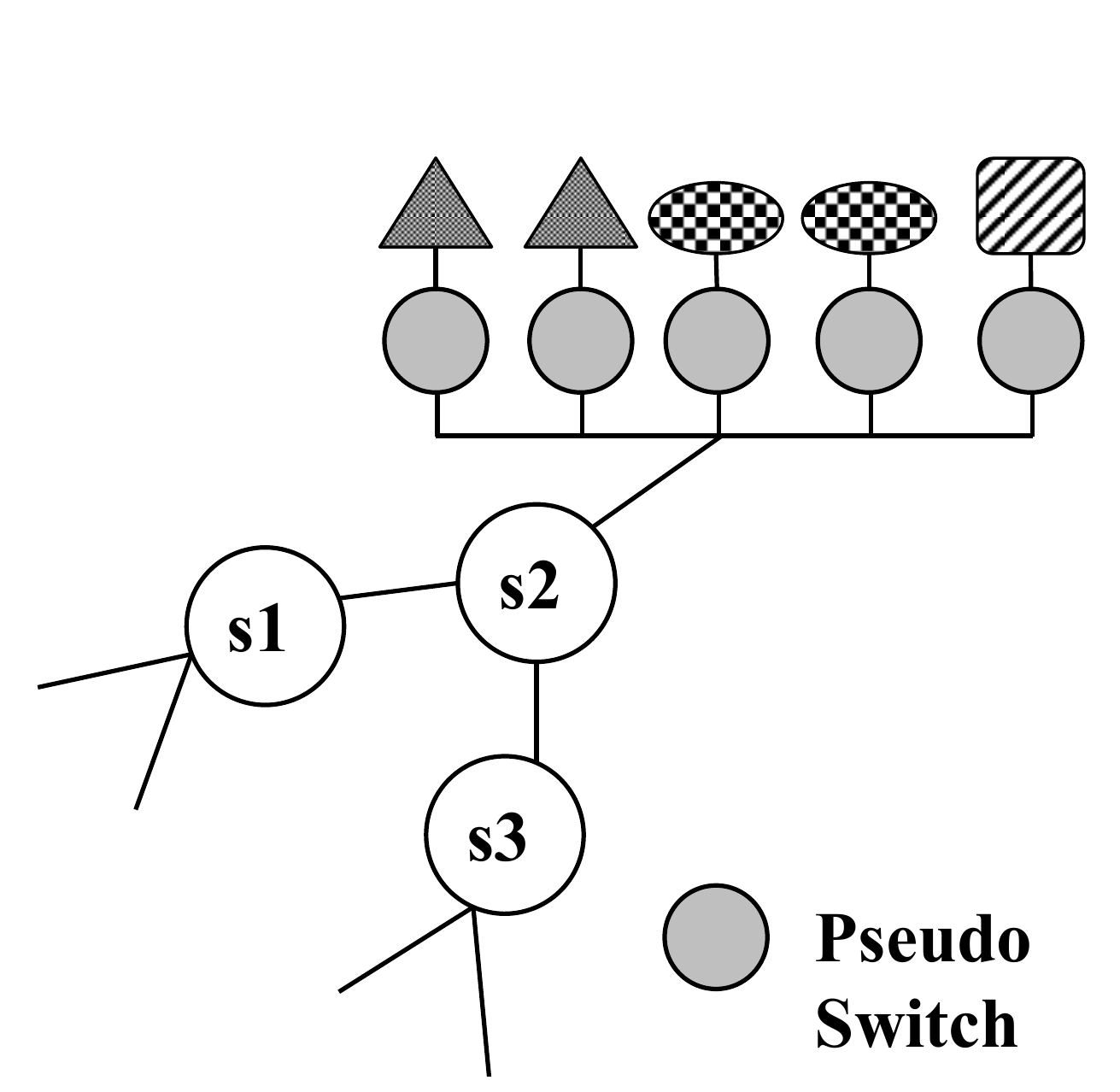}\label{fig:topology_augmentation}}
	\caption{Network Transformation}
	\label{fig:transformation}
\end{figure*}

\subsubsection{VNF Enumeration}
A part of the original physical network topology is shown in~\fig{fig:initial_network}. Here, we have three switches ($s1, s2$ and $s3$) and a server $n2$ connected to switch $s2$. The first transformation is called VNF enumeration, as we enumerate all possible VNFs in this step. The modified network after the first transformation is shown in \fig{fig:vnf_enumeration}. In this step, we find the maximum number for each type of VNF that can be deployed on each server. We calculate this number based on the resource capacity of the server and the resource requirement of a type of VNF. For example, if a server has 16 cores, and CPU requirement for Firewall and IDS are 4 and 8 cores, respectively, we can deploy 4 Firewalls and 2 IDSs on it. In~\fig{fig:vnf_enumeration} we show enumerated VNFs for server $n2$.

We denote the set of these VNFs (called pseudo-VNFs) by $\mathcal{M}$. Each VNF $m \in \mathcal{M}$ is implicitly attached to a server $\bar{n} \in \bar{N}$. We use the function $\zeta(m)$ to denote this mapping. 

\begin{align*}
\zeta(m) = \bar{n} \text{ if VNF } m \text{ is attached to server } \bar{n}
\end{align*}

We also define a function $\Omega(\bar{n})$ to represent this mapping in the opposite direction: 

\begin{align*}
\Omega(\bar{n}) = \{ m\ |\ \zeta(m) = \bar{n} \},\ m \in \mathcal{M}, \bar{n} \in \bar{N}
\end{align*}

Next, we define $q_{mp} \in \{0, 1\}$ to indicate the type of a VNF:

\begin{align*}
q_{mp} & =\left\{ \begin{array}{rl}
1 & \text{if VNF } m \text{ is of type } p \in P,\\
0 & \text{otherwise}.
\end{array}\right.
\end{align*}

As discussed earlier, a given type of VNF can be deployed on a specific set of servers. To ensure this we must have:

\begin{align}
q_{mp} = d_{\zeta(m)p}
\end{align}

We should note that pseudo-VNFs simply represent where a particular type of VNF can be provisioned. $y_m \in \{0, 1\}$ indicates whether a pseudo-VNF is active or not.

\begin{align*}
y_{m} & =\left\{ \begin{array}{rl}
1 & \text{if pseudo-VNF } m \in \mathcal{M} \text{ is active},\\
0 & \text{otherwise}.
\end{array}\right.
\end{align*}

\subsubsection{Adding Pseudo-Switches}
Next, we augment the physical topology again by adding a pseudo-switch between each pseudo-VNF and the original switch to which it was connected. This process is shown in~\fig{fig:topology_augmentation}. We perform this step to simplify the expressions of the network flow conservation constraint in the ILP formulation presented next. This process does not increase the size of the solution space as we consider them only for the flow conservation constraint. 

\subsection{ILP Formulation}\label{subsec:formulation}
We define the decision variable $x^t_{nm}$ to represent the mapping of a traffic node to a pseudo-VNF:

\begin{align*}
x_{nm}^t & =\left\{ \begin{array}{rl}
1 & \text{if node } n \in N^t \text{ is provisioned on } m \in \mathcal{M},\\
0 & \text{otherwise}.
\end{array}\right.
\end{align*}

Next, we define another variable to represent the mapping between a traffic node and a switch in the physical network.

\begin{align*}
z_{n\bar{s}}^t & =\left\{ \begin{array}{rl}
1 & \text{if node } n \in N^t \text{ is attached to switch } \bar{s},\\
0 & \text{otherwise}.
\end{array}\right.
\end{align*}

$z_{n\bar{s}}^t$ is not a decision variable as it can be derived from $x_{nm}^t$:

\begin{align*}
z_{n\bar{s}}^t = 1 \text{ if } x_{nm}^t = 1 \text{ and } \bar{h}_{\zeta(m)\bar{s}} = 1
\end{align*}

We can also derive the variable $y_m$ from $x_{nm}^t$ as follows:

\begin{align*}
y_{m} = 1 \text{ if } \sum\limits_{t \in T} \sum\limits_{n \in N^t} x_{nm}^t > 0
\end{align*}

We assume that $\hat{x}^t_{nm}$ represents the value of $x^t_{nm}$ at the last traffic provisioning event. To ensure that resources for previously provisioned traffic are not deallocated we must have $x^t_{nm} \ge \hat{x}^t_{nm},\ \forall\ t \in \hat{T}, n \in N^t, m \in \mathcal{M}$. Now, we define $\hat{y}_m \in \{0, 1\}$ that represents the value of $y_m$ at the last traffic provisioning event as follows:

\begin{align*}
\hat{y}_{m} = 1 \text{ if } \sum\limits_{t \in T} \sum\limits_{n \in N^t} \hat{x}_{nm}^t > 0
\end{align*}

Again, to ensure that resources for previously provisioned traffics are not deallocated we must have $y_m \ge \hat{y}_{m},\ \forall\ m \in \mathcal{M}$. Next, we need to ensure that VNF capacities are not over-committed. The processing capacity of an active VNF must be greater than or equal to the total amount of traffic passing through it. We express this constraint as follows:

\begin{align}
\sum\limits_{t \in T} \sum\limits_{n \in N^t} x_{nm}^t \times \beta^t \le c_m,\ \forall\ m \in \mathcal{M} | y_{m} = 1
\end{align}

We also need to make sure that physical server capacity constraints are not violated by the deployed VNFs. We represent this constraint as follows:

\begin{align}
\sum\limits_{m \in \Omega(\bar{n})} y_{m} \times \kappa_m^r \le c_{\bar{n}}^r,\ \forall\ \bar{n} \in \bar{N}, r \in R
\end{align}

Each node of a traffic must be mapped to a proper VNF type. This constraint is represented as follows:

\begin{align}
x^t_{nm} \times g^t_{np} = q_{mp},\ \forall\ t \in T, n \in N^t, m \in \mathcal{M}, p \in P
\end{align}

Next, we need to ensure that every traffic node is provisioned and to exactly one VNF.

\begin{align}
\sum\limits_{t \in T} \sum\limits_{n \in N^t}  x^t_{nm} = 1,\ \forall\ m \in \mathcal{M}
\end{align}

Now, we define our second decision variable to represent the mapping between links in the traffic model (\fig{fig:traffic_model}) to the links in the physical network.

\begin{align*}
w_{\bar{u}\bar{v}}^{tn_{1}n_{2}} & =\left\{ \begin{array}{rl}
1 & \text{if } (n1, n2) \in L^t \text{ uses physical link } (\bar{u}, \bar{v}),\\
0 & \text{otherwise}.
\end{array}\right.
\end{align*}

We also assume that $\hat{w}_{\bar{u}\bar{v}}^{tn_{1}n_{2}}$ represents the value of $w_{\bar{u}\bar{v}}^{tn_{1}n_{2}}$ at the last traffic provisioning event. To ensure that resources for previously provisioned traffics are not deallocated in the current iteration we must have

\begin{align}
w_{\bar{u}\bar{v}}^{tn_{1}n_{2}} \ge \hat{w}_{\bar{u}\bar{v}}^{tn_{1}n_{2}}, \forall\ t \in \hat{T}, n_1, n_2 \in N^t | n_2 \in \eta^t(n_1) \nonumber \\ \text{ and } n_2 > n_1, \bar{u}, \bar{v} \in \bar{S}
\end{align}

To ensure that each directed link in a traffic request is not mapped to both directions of a physical link, we must have:

\begin{align}
w_{\bar{u}\bar{v}}^{tn_{1}n_{2}} + w_{\bar{v}\bar{u}}^{tn_{1}n_{2}} \le 1, \forall\ t \in T, \nonumber \\ n_1, n_2 \in N^t | n_2 \in \eta^t(n_1) \text{ and } n_2 > n_1, \bar{u}, \bar{v} \in \bar{S}
\end{align}

Now, we present the capacity constraint for physical links:

\begin{align}
\sum\limits_{\bar{u} \in \bar{S}} \sum\limits_{\bar{v} \in \bar{S}} (w_{\bar{u}\bar{v}}^{tn_{1}n_{2}} + w_{\bar{v}\bar{u}}^{tn_{1}n_{2}}) \times \beta^t \le \beta_{\bar{u}\bar{u}}, \nonumber \\ \forall\ t \in T, n_1, n_2 \in N^t | n_2 \in \eta^t(n_1) \text{ and } n_2 > n_1
\end{align}

Next, we present the flow constraint that makes sure that the in-flow and out-flow of each switch in the physical network is equal except at the ingress and egress switches:

\begin{align}
\sum\limits_{\bar{v} \in \eta(\bar{u})} \Big(  w_{\bar{u}\bar{v}}^{tn_{1}n_{2}} - w_{\bar{v}\bar{u}}^{tn_{1}n_{2}} \Big) = z_{n_1\bar{u}}^t -   z_{n_2\bar{u}}^t, \nonumber \\ \forall\ t \in T, n_1, n_2 \in N^t | n_2 \in \eta^t(n_1) \text{ and } n_2 > n_1, \bar{u} \in \bar{S} 
\end{align}

Finally, we need to ensure that every link in a traffic request is provisioned on one or more physical links in the network:

\begin{align}
\sum\limits_{\bar{u} \in \bar{S}} \sum\limits_{\bar{v} \in \bar{S}} (w_{\bar{u}\bar{v}}^{tn_{1}n_{2}} + w_{\bar{v}\bar{u}}^{tn_{1}n_{2}}) \ge 0, \nonumber \\ \forall\ t \in T, n_1, n_2 \in N^t | n_2 \in \eta^t(n_1) \text{ and } n_2 > n_1
\end{align}

Our objective is to find the optimal number and placement of VNFs that minimizes OPEX and physical resource fragmentation in the network. We formulate them in detail below:

\textbf{-- OPEX: } We consider four cost components to contribute to OPEX. These are as follows:

\emph{1. VNF Deployment Cost:}  the VNF deployment cost can be expressed as follows:

\begin{align}\label{eqn:deployment_cost}
\mathbb{D} = \sum\limits_{m \in \mathcal{M}|y_m = 1} \mathcal{D}_{p}^{+} \times q_{mp} \times (y_{m} - \hat{y}_{m})
\end{align}

\emph{2. Energy Cost:} Without loss of generality we assume that the energy consumption of a server is proportional to the amount of resources being used. However, a server usually consumes power even in the idle state. So, we compute the power consumption of a server as follows:

\begin{align*}\label{eqn:energy_cost}
\mathbb{E}_{\bar{n}} = \sum\limits_{m \in \Omega{\bar{n}}} y_m \times q_{mp} \times e^r(c_{\bar{n}}^r, \kappa_p^r)
\end{align*}

where

\begin{align*}
e^r(r_t, r_c) = (e^r_{max} - e^r_{idle}) \times \frac{r_c}{r_t} + e^r_{idle}
\end{align*}

Here, $r_t$ and $r_c$ denote the total and consumed resource, respectively. $e^r_{idle}$ and $e^r_{max}$ denote the energy cost in the idle and peak consumption state for resource $r$, respectively.  

Now, the total energy cost is

\begin{align}
\mathbb{E} = \sum\limits_{\bar{n} \in \bar{N}} \sum\limits_{m \in \Omega{\bar{n}}} y_m \times q_{mp} \times e^r(c_{\bar{n}}^r, \kappa_p^r)
\end{align}

\emph{3. Cost of Forwarding Traffic:}
Let us assume that the cost of forwarding 1 Mbit data through one link in the network is $\sigma$ (in dollars). Now, we can compute the total cost of traffic forwarding as follows:

\begin{align}\label{eqn:traffic_forwarding_cost}
\mathbb{F} = \sum\limits_{t \in T} \sum\limits_{n_1 \in N^t} \sum\limits_{\substack{n_2 \in \eta^t(n_1)\\\text{ and }n_2 > n_1}} \sum\limits_{\bar{u} \in \bar{S}} \sum\limits_{\bar{v} \in \eta(\bar{u})} \Big( (w_{\bar{u}\bar{v}}^{tn_{1}n_{2}} - \nonumber \\ \hat{w}_{\bar{u}\bar{v}}^{tn_{1}n_{2}})  \times \beta^t \times \sigma \Big)
\end{align}

\emph{4. Penalty for SLO violation:} We can compute the actual propagation delay experienced by a traffic as follows:

\begin{align*}
\delta_t^a =  \sum\limits_{n_1 \in N^t} \sum\limits_{\substack{n_2 \in \eta^t(n_1)\\\text{ and }n_2 > n_1}} \sum\limits_{\bar{u} \in \bar{S}} \sum\limits_{\bar{v} \in \eta(\bar{u})} w_{\bar{u}\bar{v}}^{tn_{1}n_{2}} \delta_{\bar{u}\bar{v}}
\end{align*}

Let $\rho^t(\omega^t, \delta^t, \delta^t_a)$ be a function that computes the penalty for SLO violation given the policy for determining penalty ($\omega^t$), expected propagation delay ($\delta^t$) and actual propagation delay ($\delta^t_a$) for traffic $t$. So, the total cost for SLO violations can be expressed as follows:

\begin{align}\label{eqn:sla_violation_cost}
\mathbb{P} = \sum\limits_{t \in T} \rho^t(\omega^t, \delta_t, \delta^t_a)
\end{align}

\textbf{-- Resource Fragmentation:} Our second objective is to minimize resource (\eg\ server and links) fragmentation. We represent it in terms of dollar as we do for the above mentioned costs. For this purpose, we assume that $p^r$ denotes the price of unit resource of type $r \in R$. We also denote $p^{\beta}$ as the price of unit bandwidth. Now, we can compute the total cost for resource fragmentation as follows:

\begin{align}
\mathbb{C} = \sum\limits_{\bar{n} \in \bar{N}} \sum\limits_{r \in R} \Big ( c^r_{\bar{n}} - \sum\limits_{m \in \omega(\bar{n})} (\kappa^r_p \times q_{mp} \times y_m) \Big ) \times p^r + \nonumber \\ \sum\limits_{\bar{u} \in \bar{S}} \sum\limits_{\bar{v} \in \eta(\bar{u})} \Big ( \beta_{\bar{u}\bar{v}} - \sum\limits_{t \in T} \sum\limits_{n_1 \in N^t} \sum\limits_{\substack{n_2 \in \eta^t(n_1)\\\text{ and }n_2 > n_1}}  (w_{\bar{u}\bar{v}}^{tn_{1}n_{2}} \times \beta^t) \Big ) \times p^\beta \nonumber
\end{align}

Here, the first term represents the cost of server resource fragmentation (\eg\ CPU, memory, disk, \etc) and the second term represents the cost of link bandwidth fragmentation.

Our objective is to minimize the total network operational cost and resource fragmentation that can be expressed as a weighted sum of the aforementioned costs.

\begin{align}\label{eqn:objective}
\text{minimize } \Big( \alpha \mathbb{D} + \beta \mathbb{E} + \gamma \mathbb{F} + \lambda \mathbb{P} + \mu \mathbb{C} \Big) 
\end{align}

Here, $\alpha$, $\beta$, $\gamma$, $\lambda$ and $\mu$ are weighting factors that are used to adjust the relative importance of the cost components. 

The VNF-OP is NP-Hard as we can reduce this problem to the \textit{Multi-Commodity, Multi-Plant, Capacitated Facility Location Problem}~\cite{Pirkul:1998:MMC:300685.300697} or more commonly known as the \textit{Trans-shipment Problem}~\cite{chiou2008transshipment} by imposing a constraint on the maximum number of VNFs that can be deployed in the network. Both of these problems are known to be NP-Hard. So, VNF-OP is NP-Hard as well. Therefore in the next section we propose a heuristic to solve this problem.
\section{Heuristic Solution}\label{sec:heuristic}
In this section, we present a heuristic to solve the VNF-OP. Given a network topology, a set of middlebox specifications and a batch of traffic requests, the heuristic finds the number and locations of different types of VNFs required to operate the network with minimal OPEX. We did not explicitly consider resource fragmentation to keep the heuristic simple and fast. However, our experimental results show that even with this simplification, the heuristic produces solutions that are very close to the optimal. The heuristic runs in two steps. First, we model the VNF-OP as a multi-stage directed graph  with associated costs. Then we find a near-optimal VNF placement from the multi-stage graph by running the Viterbi algorithm~\cite{forney1973viterbi}. In the following, we first describe the modeling of VNF-OP using multi-stage graph (\sect{subsec:model}), followed by  the solution using Viterbi algorithm (\sect{subsec:viterbi}). A detailed discussion of the heuristic along with an illustrative example is provided in the Appendix. 

\subsection{Modeling with Multi-Stage Graph}\label{subsec:model}
For a given traffic request, $t =  \langle \bar{u}^t, \bar{v}^t, \Psi^t, \beta^t, \delta^t, \omega^t \rangle$, we represent $t$ as a multi-stage graph with $l_{\Psi^t} + 2$ stages. The first and the last $(l_{\Psi^t} + 2)$ stages represent the ingress and egresses switches, respectively. These two stages contain only one node representing $\bar{u}^t$ and $\bar{v}^t$, respectively. Stage $i$ ($\forall i \in \{2, \ldots (l_{\Psi^t} + 1)\}$), represents the $(i-1)$-th VNF and the node(s) within this stage represent the possible server locations where a VNF can be placed. Each node is associated with a VNF deployment cost (\eqn{eqn:deployment_cost}) and an energy cost (\eqn{eqn:energy_cost}) as described in \sect{subsec:formulation}.

An edge $(\bar{v}_i, \bar{v}_j)$ in this multi-stage graph represents the placement of a VNF at a server attached to switch $\bar{v}_j$, given that the previous VNF in the sequence is deployed on a server attached to switch $\bar{v}_i$. We put a directed edge between all pairs of nodes in stage $i$ and $i + 1$ ($\forall i \in \{1, 2, \ldots (l_{\Psi^t} + 1)\}$). We associate two costs with each edge: the cost for forwarding traffic (\eqn{eqn:traffic_forwarding_cost}) and the penalty for SLO violations (\eqn{eqn:sla_violation_cost}). The traffic forwarding cost is proportional to the weighted shortest path (in terms of latency) between the switches. The penalty for SLO violations is obtained by the following process: (i) we equally divide the maximum allowed delay between the stages, (ii) we assign a SLO violation cost for a transition between two successive stages in the multi-stage graph whenever we incur more than the allocated delay due to traffic transport and processing at the nodes. The total cost of a transition between two successive stages in the multi-stage graph is calculated by summing the node and edge costs following~\eqn{eqn:objective}. Finally, a path from the node in the first stage to the node in the last stage represents a placement of the VNFs. Our goal is to find a path in the multi-stage graph that yields minimal OPEX.

\begin{algorithm}[b]
	\small
	\caption{ProvisionTraffic($t$, $\bar{G}$)}\label{alg:place-middlebox}
	\begin{algorithmic}[1]
		\STATE $\forall (i,j) \in \{1 \ldots |\Psi^t|\} \times \{1 \ldots| \bar{S}|\}$ : $cost_{i,j} \leftarrow \infty$, $\pi_{i,j} \leftarrow NIL$
		\STATE $\forall i \in |\bar{S}|$ :
		\IF{$IsResourceAvailable(u^t, i, \Psi^t_1, t)$}
		\STATE $cost_{1,n} \leftarrow GetCost(u^t, i, \Psi^t_1, t)$, $\pi_{1,n} \leftarrow n$
		\ENDIF
		\STATE $\forall (i,j,k) \in \{2 \ldots |\Psi^t|\} \times \{1 \ldots| \bar{S}| \times \{1 \ldots| \bar{S}|\}$ :
		\IF{$IsResourceAvailable(k,j,\Psi^t_i,t)$}
		\STATE $cost_{i,j} \leftarrow$ $\min\{cost_{i,j}, cost_{i-1,k} + GetCost(k,j,\Psi^t_i,t)\}$
		\STATE $\pi_{i,j} \leftarrow$ \text{$i$ yielding minimum $cost_{i,j}$} 
		\ENDIF
		\STATE $\Pi \leftarrow NIL$, $C \leftarrow \infty$, $\psi \leftarrow <>$
		\STATE $\forall i \in |\bar{S}|$ :
		\STATE \hspace{0.1in} $C \leftarrow \min\{C, cost_{|\Psi^t|,i} + ForwardingCost(i,v^t) + $\\ \hspace{0.2\textwidth}$SLOViolationCost(i,v^t,t)\}$			
		\STATE \hspace{0.1in} $\Pi \leftarrow$ \text{$i$ yielding minimum $cost_{|\Psi^t|,i}$}
		\STATE $\forall i \in <|\Psi^t|, |\Psi^t| - 1 \ldots 1>$ : \text{Append} $\Pi$ to $\psi$, $\Pi \leftarrow \pi_{i,\Pi}$
		\RETURN $Reverse(\psi)$
	\end{algorithmic}	
\end{algorithm}

\subsection{Finding a Near-Optimal Solution}\label{subsec:viterbi}
Viterbi algorithm a widely used method for finding the most likely sequence of states from a set of observed states. To find such a sequence, Viterbi algorithm first models the states and their relationships as a multi-stage graph. Each stage consists of the possible states and a transition cost is assigned between all pairs of states in successive stages. Once the multi-stage graph is constructed, Viterbi algorithm proceeds by computing a per node cumulative cost, $cost_u$, recursively defined as  the minimum of $cost_v + transition\_cost(v,u)$ , for all $v$ in the previous stage as of $u$'s stage. $cost_u$ represents the cost of including node $u$ in the final solution. This computation proceeds in the increasing order of stages. After finishing the computation at the final stage, the most likely sequence of states is constructed by tracing back a path from the final stage back to the first that yields the minimum cost.

We borrow the idea of how costs are computed from Viterbi Algorithm and propose a traffic provisioning algorithm, $ProvisionTraffic$ (\alg{alg:place-middlebox}). It takes a traffic request $t$ and a network topology $\bar{G}$ as input and returns a placement of $\Psi^t$ in $\bar{G}$. For each node $u$ in each stage $i$, we find a node $v$ in stage $i - 1$ that yields the minimum total cost $cost_{v, u}$ (costs are defined according to the discussion in \sect{subsec:model}). We keep track of the minimum cost path using the table $\pi$. After finishing computation for the final stage, we construct the desired VNF placement by back tracing pointers from the final stage of the multi-stage graph to the first stage, using the entries in $\pi$. During this process we update residual resource capacities of the servers and the residual bandwidth of the links after each path is allocated. For each traffic request the heuristic solution runs in $\Theta(n^2m)$ time, where $n$ is the number of switches in the network and $m$ is the VNF sequence length (See \appx{appsec:algorithm} for further details).

\section{Performance Evaluation}\label{sec:evaluation}
We perform trace driven simulations on real-world network topologies to gain a deeper insight, and to evaluate the effectiveness of the proposed solution. Our simulation is focused on the following aspects: (i) demonstrating the benefits of dynamic VNF orchestration over hardware middleboxes~(\sect{subsec:vnf_vs_hardware}), (ii) comparing the performance of the heuristic solution with that of the CPLEX based optimal solution~(\sect{subsec:cplex_vs_heuristic}) and (iii) Analyzing the behavior of the proposed solution for different traffic volume~(\sect{subsec:load_test}). Before presenting the results, we briefly describe the simulation setup~(\sect{subsec:exp-setup}) and the evaluation metrics~(\sect{subsec:exp-metric}). Implementations of both CPLEX and heuristic are available at \texttt{http://goo.gl/Da7EZu}.

\subsection{Simulation Setup}\label{subsec:exp-setup}

\subsubsection{Topology Dataset}
We have used a wide range of network topologies: (i) Internet2 research network ($12$ nodes, $15$ links)~\cite{internet2data}, (ii) A university data center network ($23$ nodes, $42$ links)~\cite{benson2010network} and (iii) Autonomous System $3967$ (AS-$3967$) from Rocketfuel topology dataset ($79$ nodes, $147$ links)~\cite{spring2002measuring}.

\subsubsection{Traffic Dataset}
We use both real traces and synthetically generated traffic for the evaluation. We use traffic matrix traces from~\cite{internet2data} to generate time varying traffic for the Internet2 topology. This trace contains a snapshot of a $12 \times 12$ traffic matrix and demonstrates significant variability in traffic volume. For the data center network, we use the traces available from~\cite{benson2010network}, and replay the traffic between random source-destination pairs. Finally, for the Rocketfuel topology, we generated a synthetic time-varying traffic matrix using the FNSS tool~\cite{fnss}. It follows the distribution from~\cite{nucci2005problem} and exhibits time-of-day effect.

\subsubsection{Middlebox and Cost Data}
We have generated a $3$-length middlebox sequence for each traffic based on the data provided in~\cite{sfcdraft,qazi2013simple}. We have used publicly available data sheets from manufacturers and service providers to select and infer values for server energy cost, SLO violation cost (for violating maximum latency), resource requirements for software middleboxes and their processing capacities. We also obtained energy consumption data for hardware middleboxes from a popular network equipment manufacturer.~\tab{tab:server-mbox-data} lists the parameters used for servers, VNFs and middleboxes. In the rest of this section we use the term ``middlebox'' to refer to both hardware middlebox and VNF.

\begin{table}
  \centering
  \caption{Server and Middlebox Data Used in Evaluation}
  \begin{tabular}{|c|c|c|}
  \hline
  \multicolumn{3}{|c|}{Server Data~\cite{serverpower}}\\
  \hline
  Physical CPU Cores & Idle Energy & Peak Energy\\
  \hline
  16 & $80.5$W & $2735$W\\
  \hline
  \multicolumn{3}{|c|}{Hardware Middlebox Data}\\
  \hline
  Idle Energy & Peak Energy & Processing Capacity\\
  \hline
  $1100$W & $1700$W & $40$Gbps \\
  \hline
  \multicolumn{3}{|c|}{VNF Data~\cite{vnfguide, martins2014clickos}}\\
  \hline
  Network Function & CPU Required & Processing Capacity\\
  \hline
  Firewall & $4$ & $900$Mbps\\
  \hline
  Proxy & $4$ & $900$Mbps\\
  \hline
  Nat & $2$ & $900$Mbps\\
  \hline
  IDS & $8$ & $600$Mbps\\
  \hline
  \end{tabular}
  \label{tab:server-mbox-data}
\end{table}

\begin{figure*}
	\centering
	\subfigure[Hardware vs. VNF (Internet2)]{\label{fig:cost_ratio_hw}\includegraphics[width=0.275\linewidth]{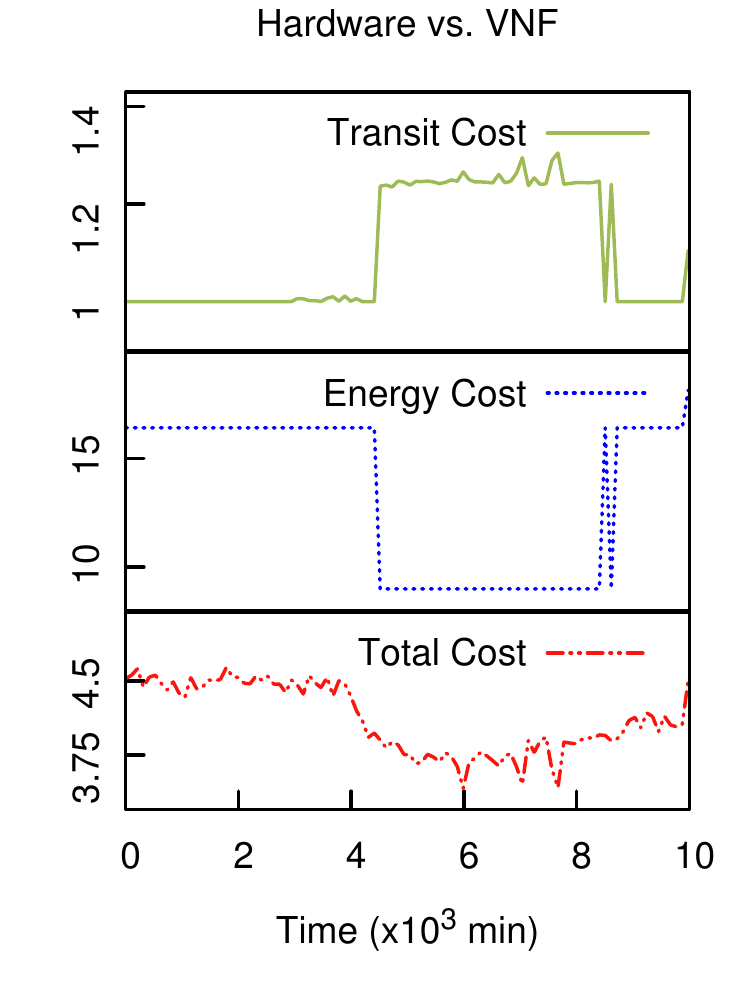}}
	\subfigure[Internet2]{\label{fig:cost_ratio_i2}\includegraphics[width=0.275\linewidth]{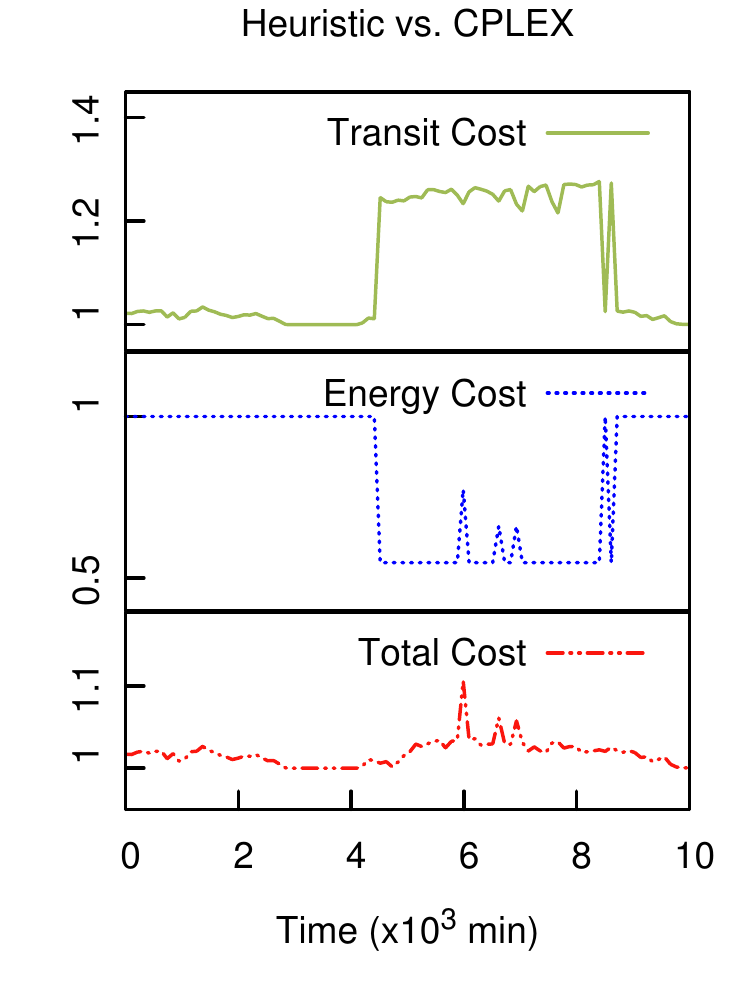}}
	\subfigure[Data Center]{\label{fig:cost_ratio_dc}\includegraphics[width=0.275\linewidth]{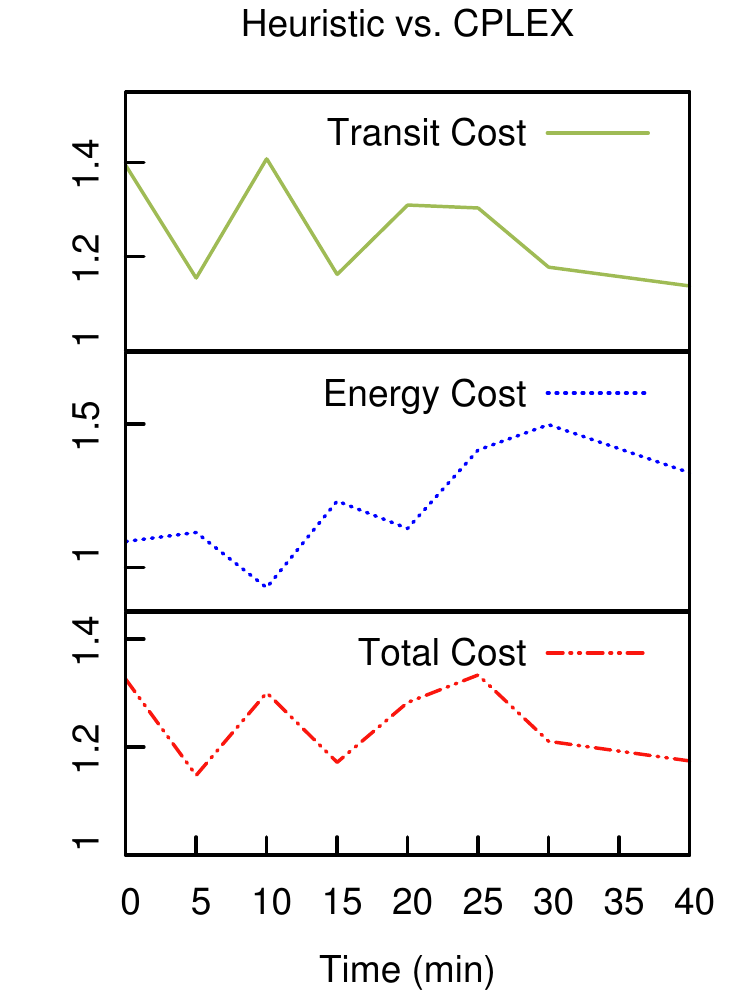}}
	\caption{Time vs. Cost Ratio}
\end{figure*}

\subsection{Evaluation Metrics}\label{subsec:exp-metric}
\subsubsection{Operational Expenditure (OPEX)} We measure OPEX according to~\eqn{eqn:objective}, and compare CPLEX and heuristic by plotting the ratio of OPEX and its components.

\subsubsection{Execution Time} It is the time required to find middlebox placement for a given traffic batch and network topology.

\subsubsection{System Utilization} We compute it as the fraction of used CPU for a server. We also report the number of active servers. 

\subsubsection{Topological Properties of Solution} We report two topological properties of the middlebox locations: (i) percentage of middleboxes placed withing $k$-hops from the ingress/egress switches and (ii) path stretch, \ie the ratio of path length obtained by CPLEX or the heuristic to the shortest path length for the traffic. The first metric gives us an insight into the location of middleboxes with respect to the ingress/egress switches, and the second one shows how many additional links (hence more bandwidth) are required to steer traffic through middlebox sequences.

\subsection{VNFs vs. Hardware Middleboxes}\label{subsec:vnf_vs_hardware}
One of the driving forces behind NFV is that VNFs can significantly reduce a network's OPEX. Here, we provide quantifiable results to validate this claim.~\fig{fig:cost_ratio_hw} shows the ratio of OPEX for hardware middleboxes to VNFs for incoming traffic provisioning requests (about $132$ requests per batch) over a period of $10000$ minutes. We show two components of OPEX: energy and transit cost. There is no publicly available data that can be used to estimate the deployment cost of hardware middleboxes. So, for this experiment, we do not consider deployment cost as a component of OPEX to make the comparison fair. The SLO violation penalty is not shown as it is zero for all time-instances. We implemented a different CPLEX program to peak provision the hardware middleboxes (peak traffic occurs at time-instance $7665$). VNFs are provisioned at each time-instance by the CPLEX program corresponding to the formulation provided in~\sect{sec:formulation}.

The bottom part of~\fig{fig:cost_ratio_hw} shows that VNFs provide more than $4\times$ reduction in OPEX. The individual reductions in energy and transit costs are also shown in the same figure. The reduction in energy cost is much higher than that of the transit cost. This is due to the fact that hardware middleboxes consume considerably higher energy than commodity servers. From~\fig{fig:cost_ratio_hw} and~\fig{fig:internet2}, we can also see that with the increase in traffic volume (after time-instance $4000$) the total cost ratio decreases. Interestingly, the energy cost ratio decreases, but the transit cost ratio increases. Handling higher traffic volume requires higher number of VNFs to be deployed, which increases the energy consumption of commodity servers, thus decreasing the energy cost ratio. However, VNFs are provisioned at optimal locations by CPLEX, which causes the transit cost to decrease and increases the transit cost ratio. The cost ratio relationship between VNFs and hardware middleboxes depends on a number of factors like processing capacity, traffic volume, idle and peak energy consumption. 

\begin{figure*}
\centering
	\subfigure[Distance between Middlebox and Ingress]{\label{sfig:ingress_hw}\includegraphics[width=0.3\textwidth]{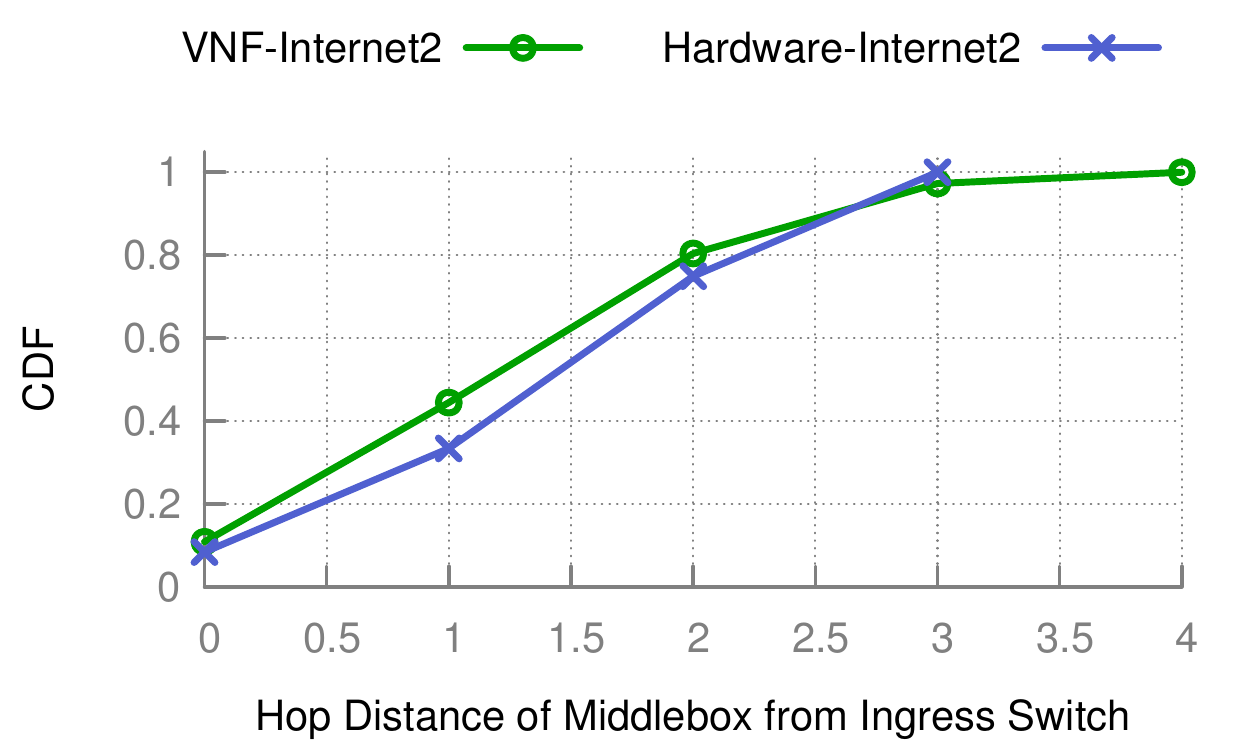}}
	\subfigure[Distance between Middlebox and Egress]{\label{sfig:egress_hw}\includegraphics[width=0.3\textwidth]{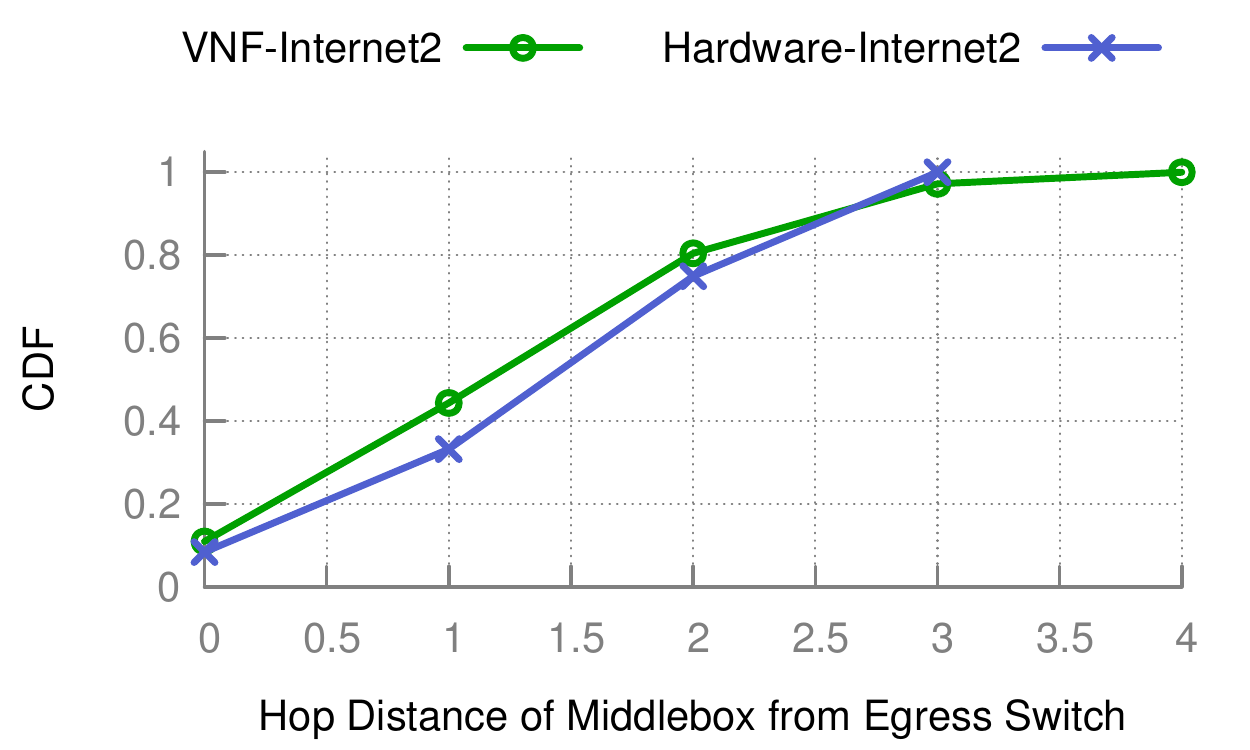}}
	\subfigure[Path Stretch]{\label{sfig:stretch_hw}\includegraphics[width=0.3\textwidth]{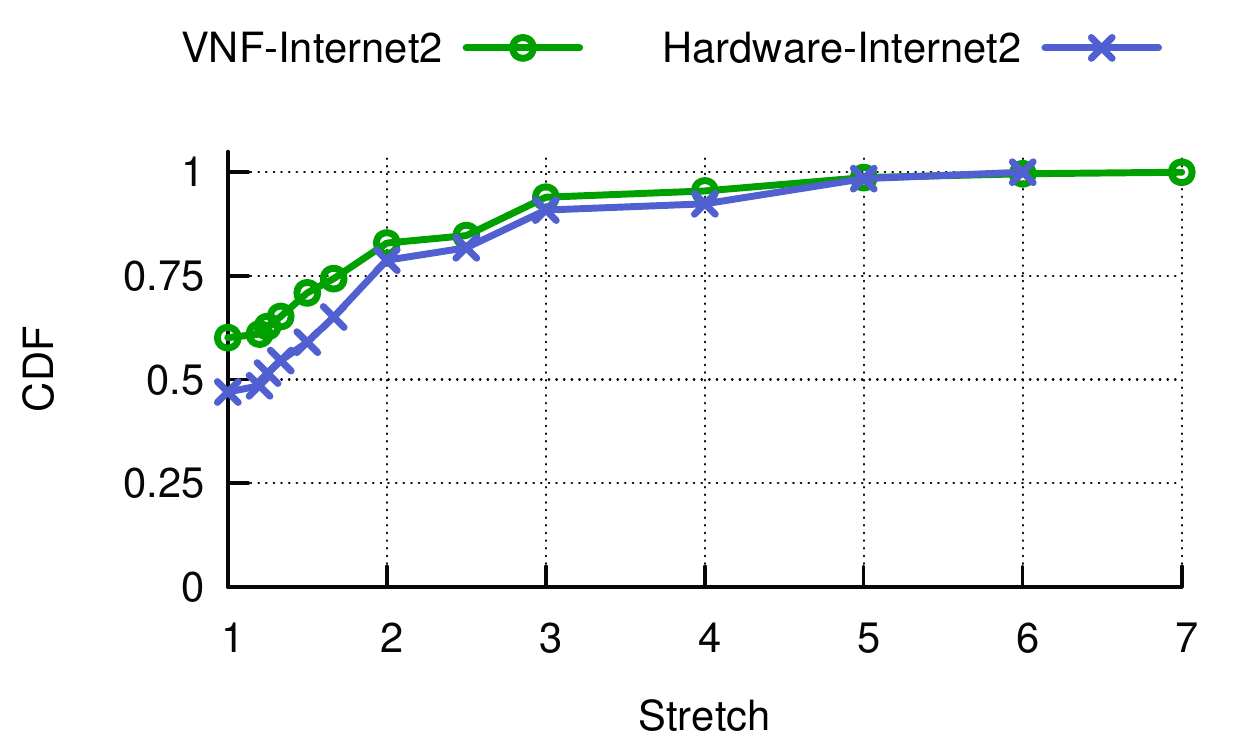}}
\caption{Topological Property Comparison between Hardware middlebox and VNF deployment (Internet2)}
\label{fig:topology_hw}
\end{figure*}

\begin{figure*}
\centering
	\subfigure[Mean Server Utilization (Internet2)]{\label{sfig:util_ts_i2}\includegraphics[width=0.3\textwidth]{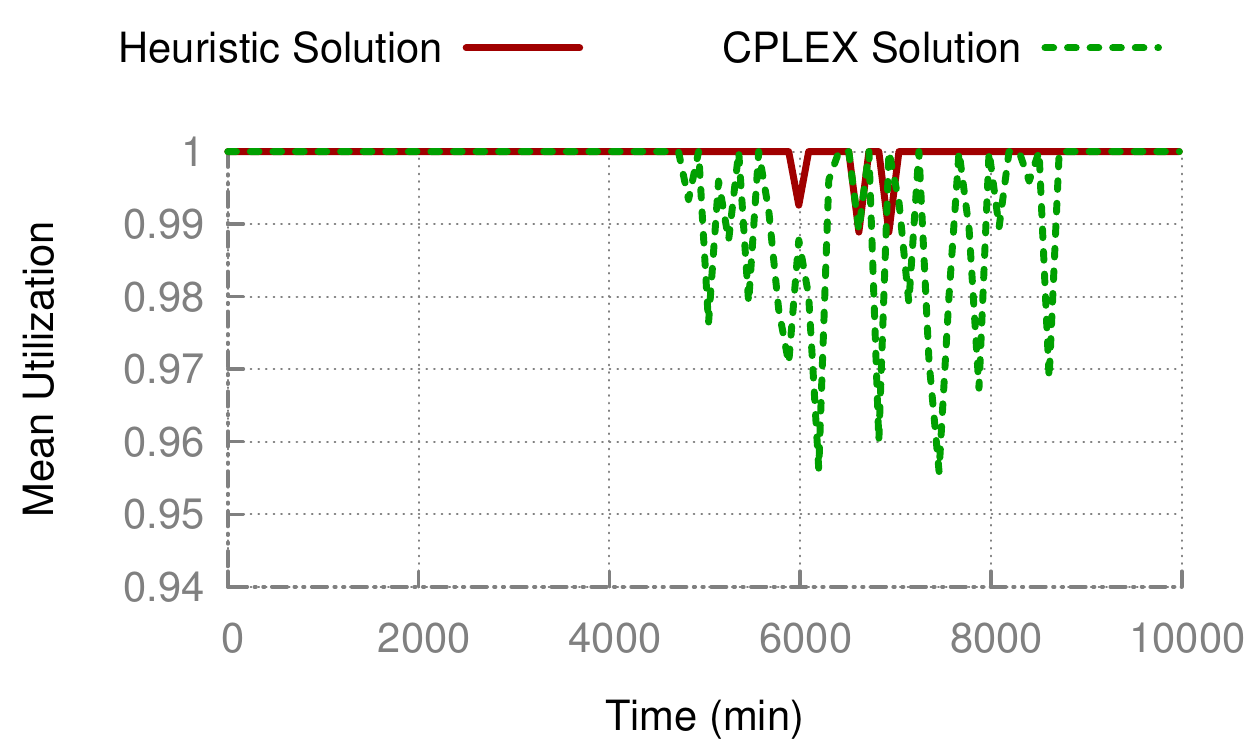}}
	\subfigure[Number of Active Servers (Internet2)]{\label{sfig:util_as_i2}\includegraphics[width=0.3\textwidth]{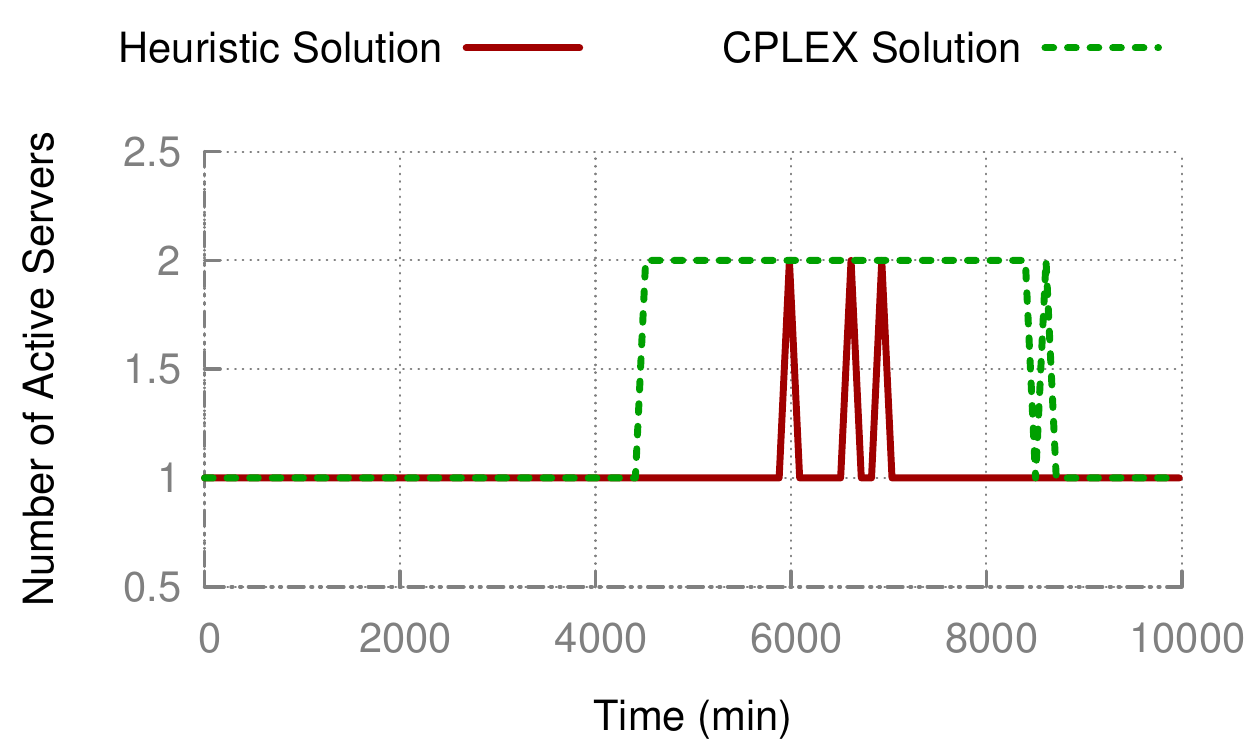}}
	\subfigure[Per Server Utilization (Internet2)]{\label{sfig:util_ps_i2}\includegraphics[width=0.3\textwidth]{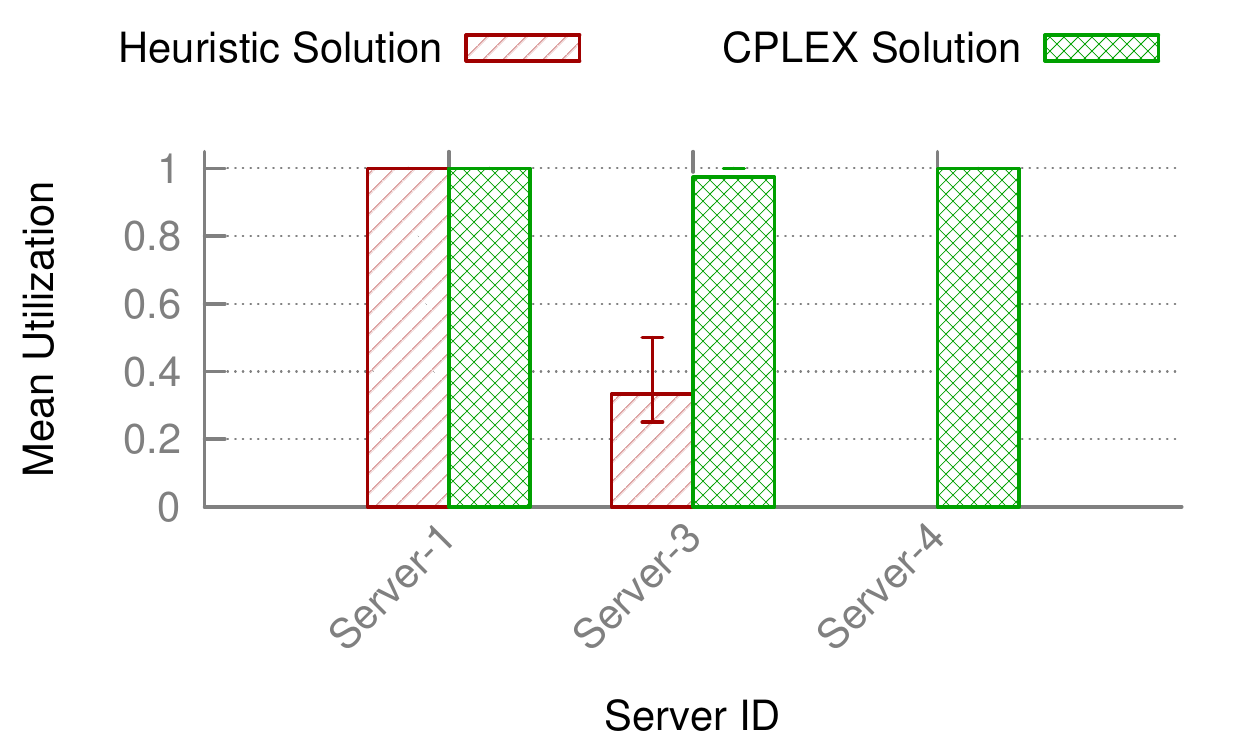}}
	\subfigure[Mean Server Utilization (Data Center)]{\label{sfig:util_ts_dc}\includegraphics[width=0.3\textwidth]{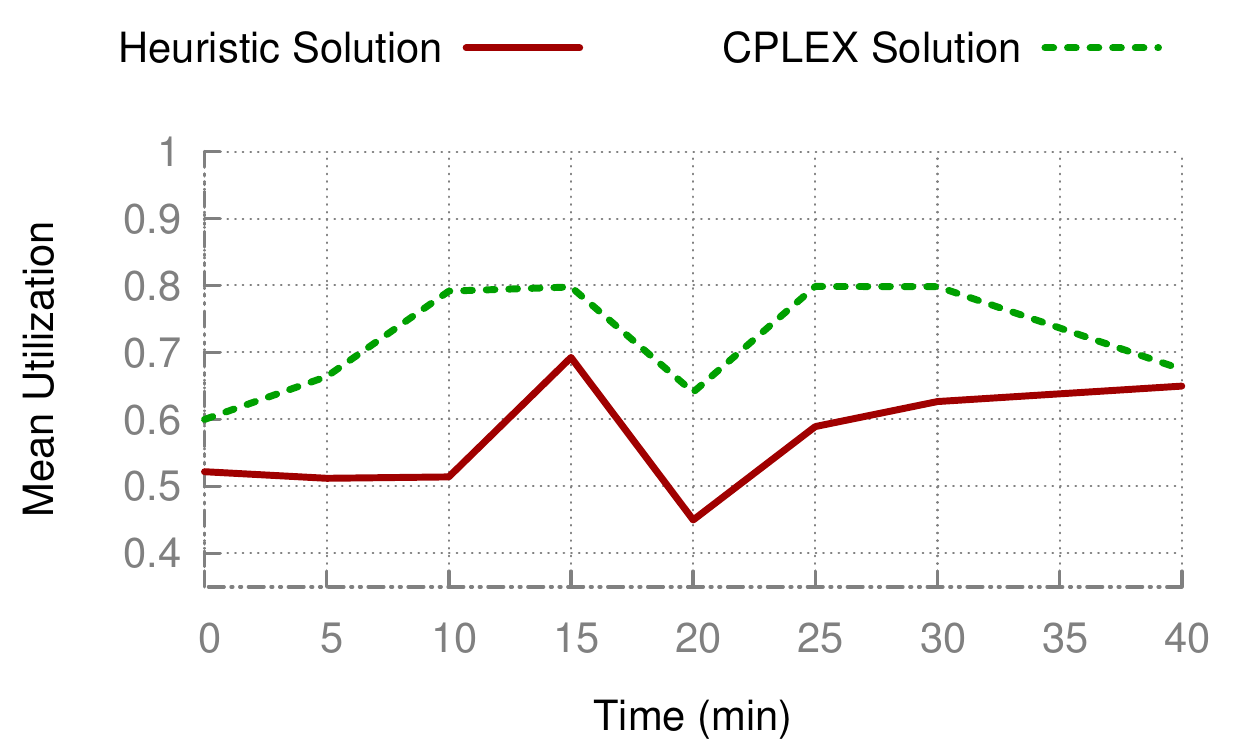}}
	\subfigure[Number of Active Servers (Data Center)]{\label{sfig:util_as_dc}\includegraphics[width=0.3\textwidth]{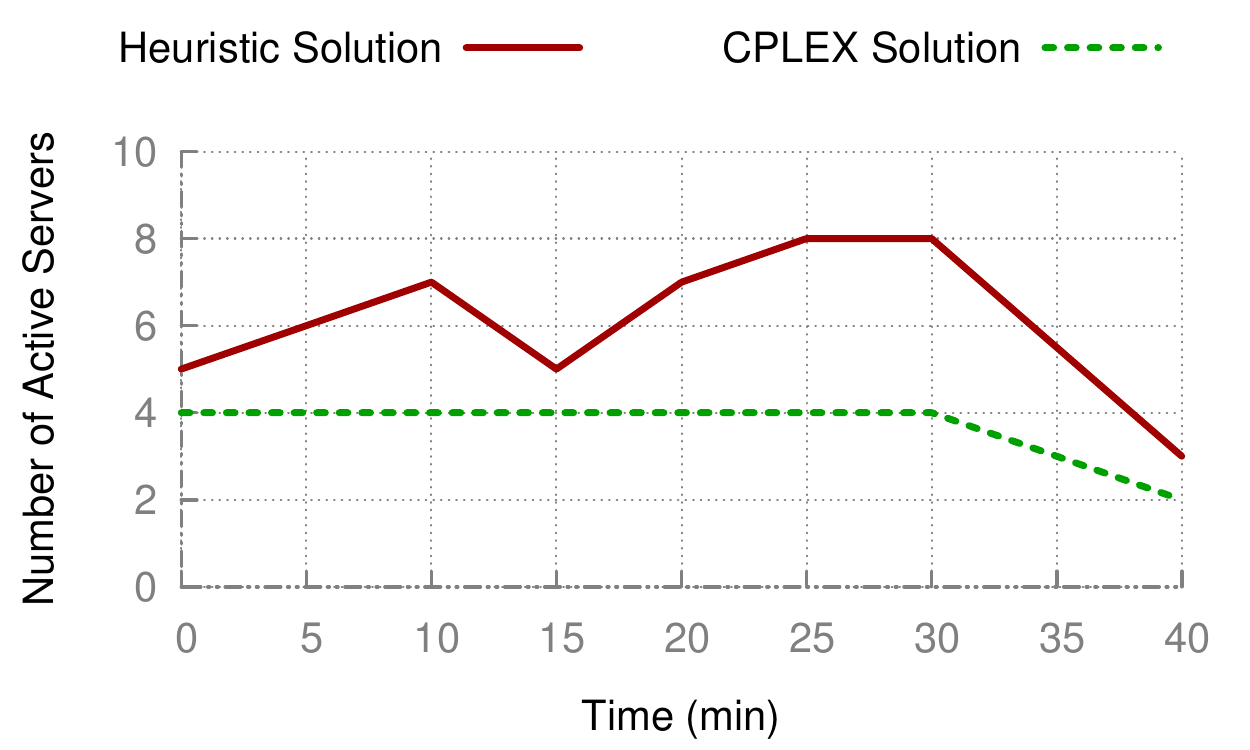}}
	\subfigure[Per Server Utilization (Data Center)]{\label{sfig:util_ps_dc}\includegraphics[width=0.3\textwidth]{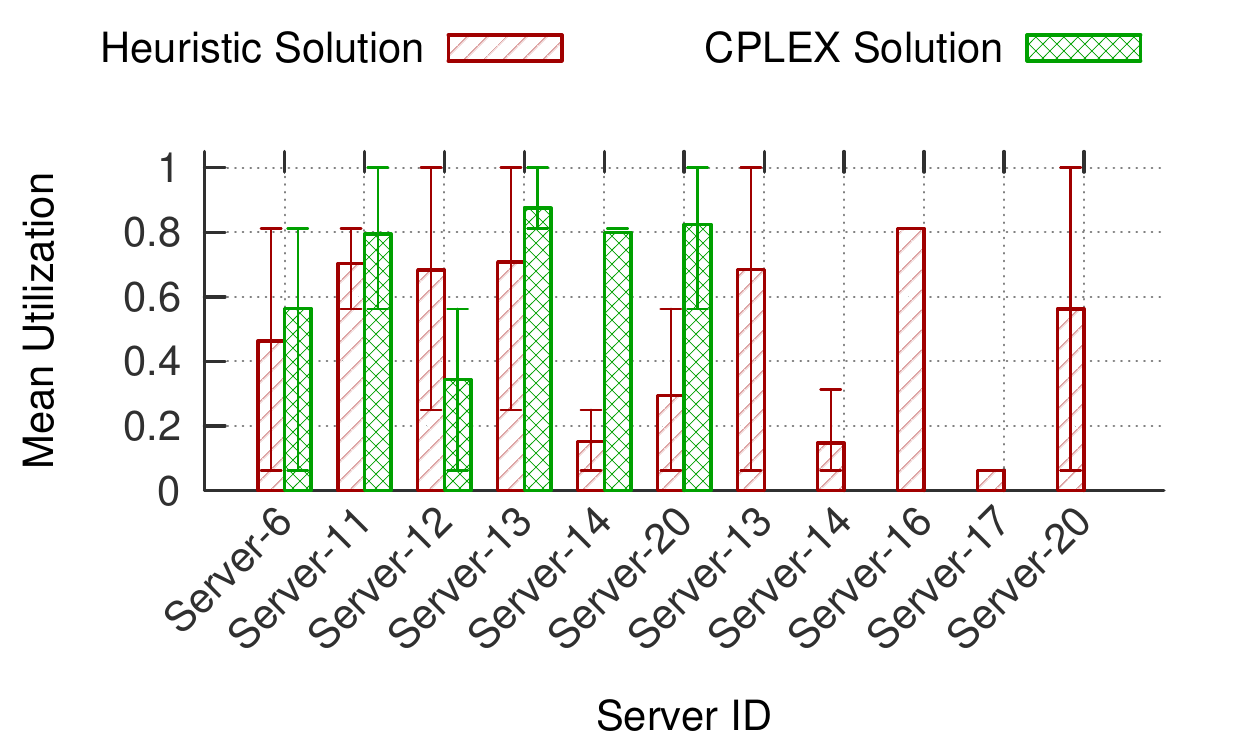}}
	\caption{Resource Utilization}
	\label{fig:resource_utilization}
\end{figure*}

\begin{figure*} [t]
	\centering
	\subfigure[Distance between middlebox and ingress]{\label{sfig:ingress}\includegraphics[width=0.3\textwidth]{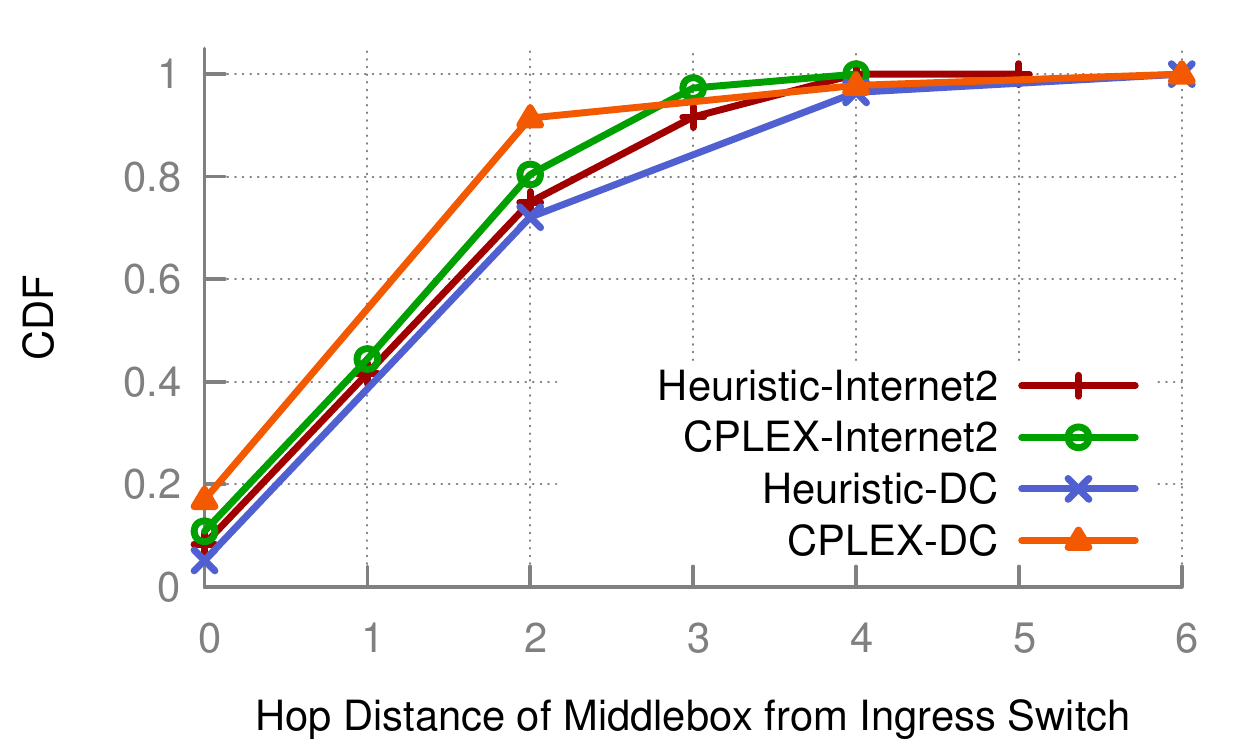}}
	\subfigure[Distance between middlebox and egress]{\label{sfig:egress}\includegraphics[width=0.3\textwidth]{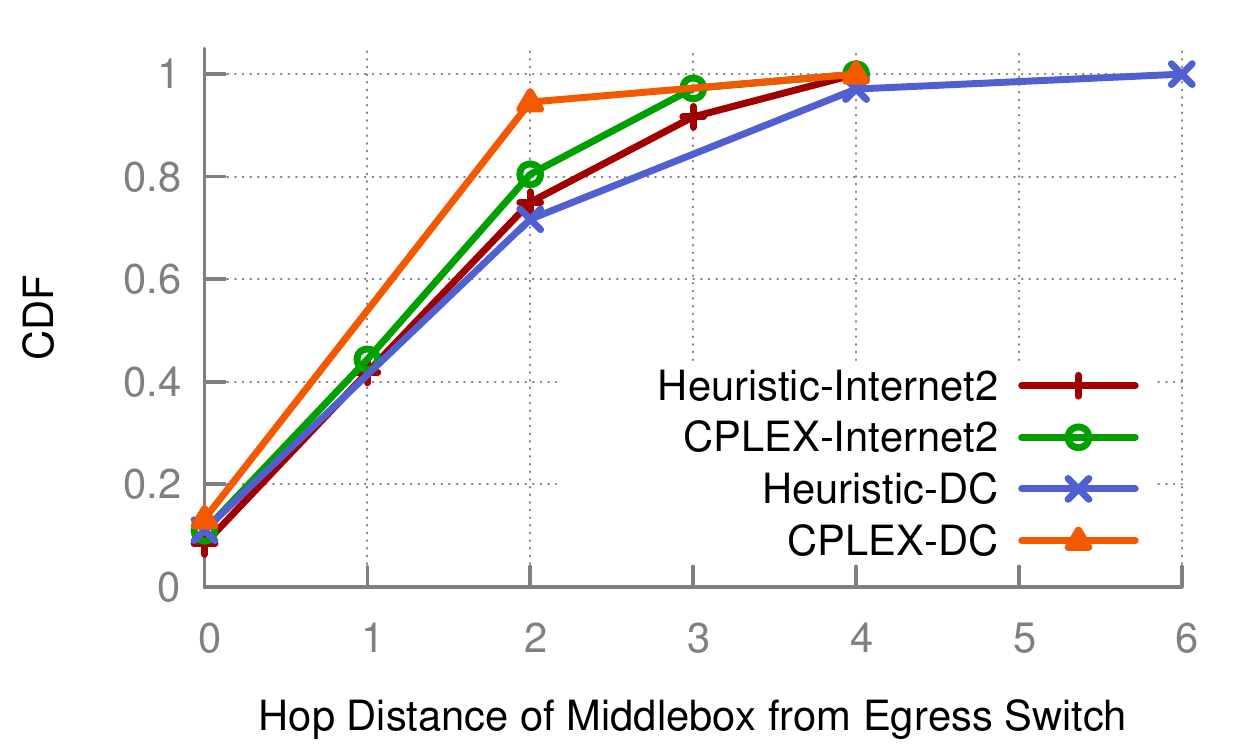}}
	\subfigure[Path stretch]{\label{sfig:stretch}\includegraphics[width=0.3\textwidth]{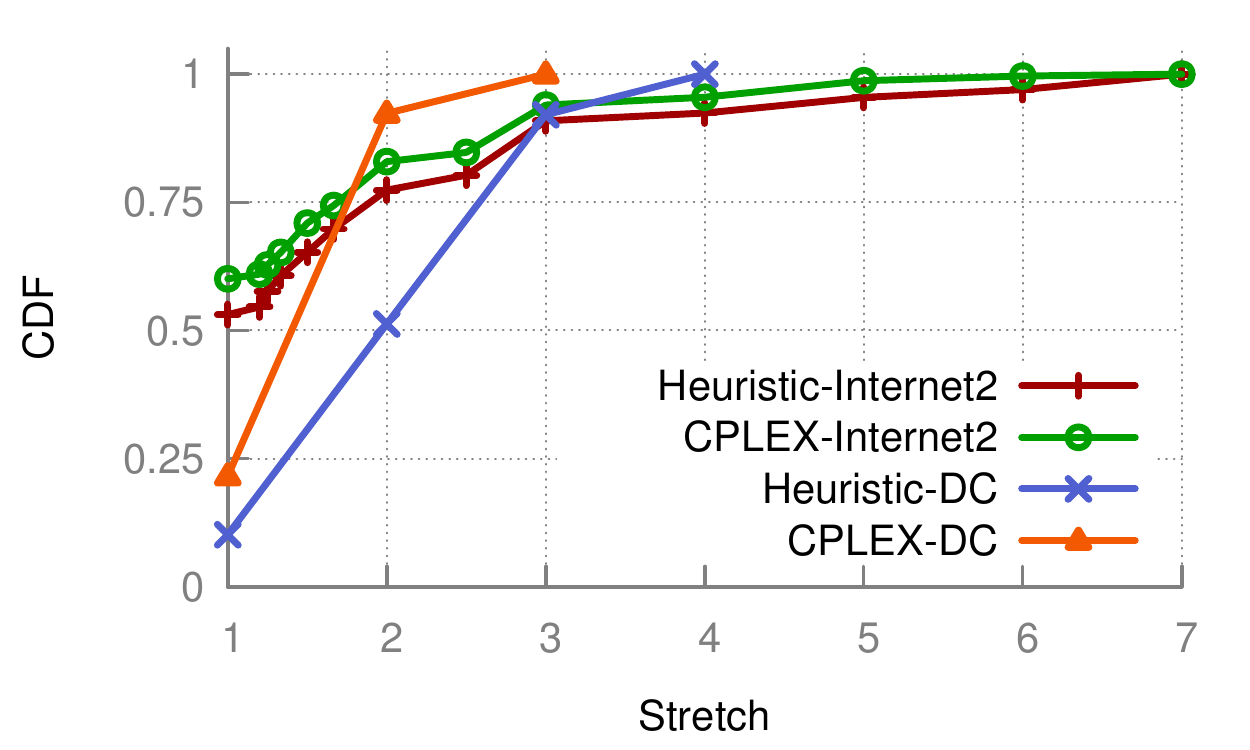}}
	\caption{Topological properties of solution}
	\label{fig:topology}
\end{figure*}

The topological properties of VNF and hardware middlebox placement locations are reported in~\fig{fig:topology_hw}. The CDF of hop distance between the ingress switch and middlebox is shown in~\fig{sfig:ingress_hw}. Higher percentage of VNFs are located within $2$ hops of the ingress switch (mostly withing $1$ hop), compared to hardware middleboxes. Some VNFs are also located at $4$ hop distance. This only occurs when placing a VNF farther away reduces the OPEX by decreasing the energy cost. Similar results are obtained for the hop distance between middlebox and egress switch~(\fig{sfig:egress_hw}). These two figures also demonstrate the fact that CPLEX places middleboxes in a more balanced (symmetric) way on the path between the ingress and egress switch. The path stretch for both hardware middleboxes and VNFs are shown in~\fig{sfig:stretch_hw}. VNFs consistently achieve a lower path stretch than hardware middleboxes, as VNF locations are not static like the hardware middleboxes. They can be provisioned on any server to reduce OPEX.

\subsection{Performance Comparison Between CPLEX and Heuristic}\label{subsec:cplex_vs_heuristic}
Now, we compare the performance of our heuristic with that of the optimal solution.~\fig{fig:cost_ratio_i2} and~\fig{fig:cost_ratio_dc} show the cost ratios for Internet2 and data center networks, respectively. The traffic patterns for these two topologies are shown in~\fig{fig:internet2} and~\fig{fig:dc}, respectively. The deployment cost and penalty for SLO violation are not shown, as the deployment cost is equal in both cases and the SLO violation penalty is zero for all time-instances. From~\fig{fig:cost_ratio_i2}, we can see that the heuristic finds solutions that are within $1.1$ times of the optimal solution. During peak traffic periods, the ratio of energy cost goes below $1$, but the ratio of transit cost increases. The optimal solution adapts to high traffic volumes by deploying more VNFs (increasing energy cost) and placing them at locations that decrease the transit cost. As a result, the ratio of energy cost decreases and the ratio of transit cost increases. However, the total cost ratio stays almost the same (varying between $1$ and $1.1$). Similar results are obtained for the data center network (\fig{fig:cost_ratio_dc}) as well. Here, the cost ratio is also very close to $1$ and varies between $1.1$ and $1.3$. 

The average execution times of the heuristic and CPLEX are shown in~\tab{tab:sol_time}. They were run on a machine with $10 \times 16$-Core $2.40$GHz Intel Xeon E7-8870 CPUs and $1$TB memory. As we can see, our heuristic provides solutions that are very close to the optimal one and its execution time is several order of magnitude faster than CPLEX.

\begin{table}
  \caption{Average Execution Time}
  \label{tab:sol_time}
  \centering
  \begin{tabular}{|c|c|c|}
    \hline
        Topology     &  CPLEX  & Heuristic\\
    \hline
    Internet2 ($12$ nodes, $15$ links) & $34.99$s & $0.535$s\\
    \hline
    Data Center ($23$ nodes, $43$ links) & $1595.12$s & $0.442$s\\
    \hline
    AS-3967 ($79$ nodes, $147$ links) & $\infty$ & $2.54$s\\
    \hline    
  \end{tabular}
\end{table}

~\fig{fig:resource_utilization} shows results related to server resource utilization for Internet2 and data center networks.~\fig{sfig:util_ts_i2} and~\fig{sfig:util_as_i2} show the mean utilization and the total number of active servers, respectively, for the Internet2 topology.~\fig{sfig:util_ps_i2} shows the average utilization per server over all time-instances. The mean utilization of the heuristic is less than that of CPLEX, as CPLEX uses more servers than the heuristic (\fig{sfig:util_as_i2}). CPLEX achieves lower OPEX by deploying more VNFs during higher traffic periods to route traffic through shorter paths. However, the solutions provided by the heuristic are within $1.1$ times the optimal one (\fig{fig:cost_ratio_i2}). In case of the data center network, CPLEX uses less servers than the heuristic (\fig{sfig:util_as_dc}) and the utilization is also higher (\fig{sfig:util_ts_dc}). The solution provided by the heuristic has higher resource fragmentation than the CPLEX one (\fig{sfig:util_ps_dc}). The data center topology offers higher number of locations to deploy VNFs compared to Internet2. Hence, the heuristic falls a little short of the optimal placement as it explores a smaller solution space. CPLEX finds the optimal value, but at the cost of much higher execution time~(\tab{tab:sol_time}). 

The topological properties for middlebox deployment for Internet2 and data center networks are shown in~\fig{fig:topology}. The CDF of hop distance from the ingress switch to a VNF is shown in~\fig{sfig:ingress}. The hop distances for the heuristic is quite close to that of the optimal solution. In case of the data center network, there is a relatively larger gap. This occurs due to the higher path diversity offered by a data center network. Each pair of nodes has more than one equal cost path. CPLEX finds the optimal solution by exploring all of them. The heuristic always picks the first shortest path. It does not explore the alternate paths to keep the execution time within practical limits (\tab{tab:sol_time}). Similar results are observed for the egress case (\fig{sfig:egress}). From~\fig{sfig:ingress} and~\fig{sfig:egress} we can also see that the CDFs are quite similar, which means that both CPLEX and heuristic place VNFs uniformly on the path between the ingress and egress switches. The path stretch is shown in~\fig{sfig:stretch}. As before, the heuristic's performance is close to that of the optimal solution. In case of the data center network, the heuristic has a larger stretch, which is a result of the path diversity issue discussed earlier.

We obtained similar results for the Rocketfuel topology. Due to space limitations they are discussed in the Appendix.

\begin{figure}[htb]
	\centering
	\subfigure[Cost Ratio]{\label{fig:cost_load_variation}\includegraphics[width=0.2405\textwidth]{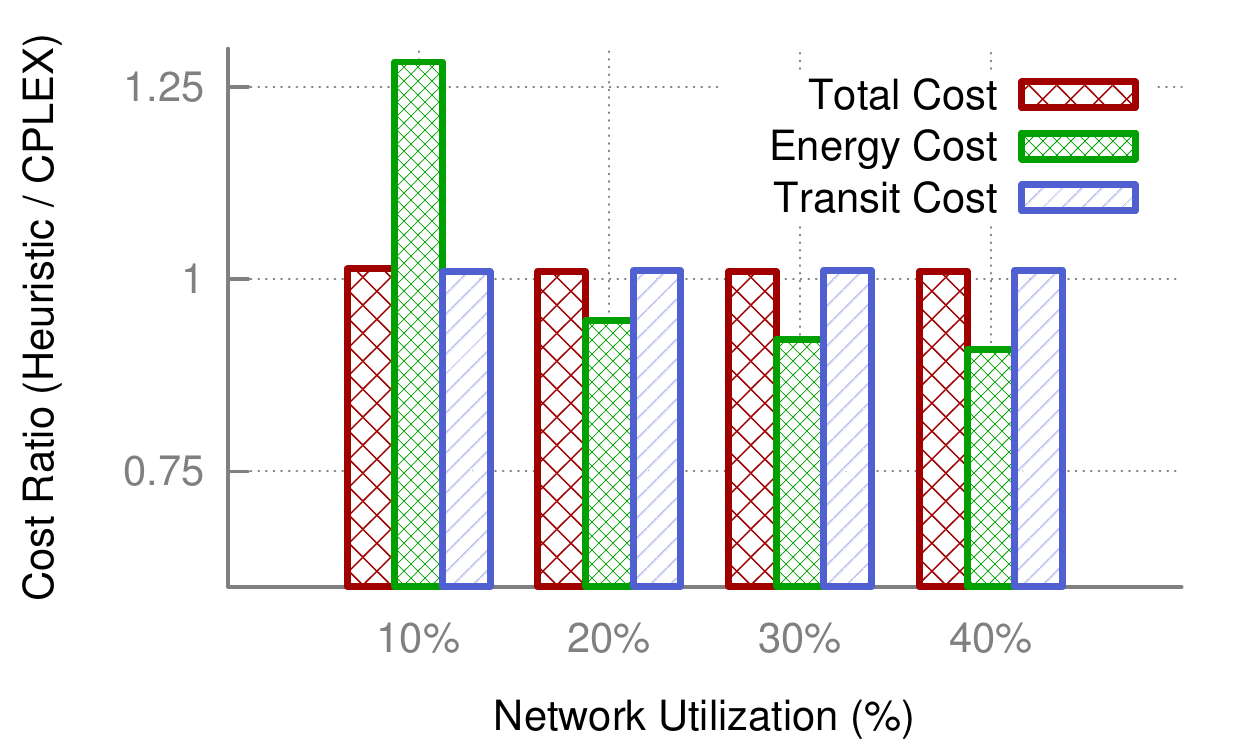}}
	\subfigure[Server Utilization]{\label{fig:util_load_variation}\includegraphics[width=0.2405\textwidth]{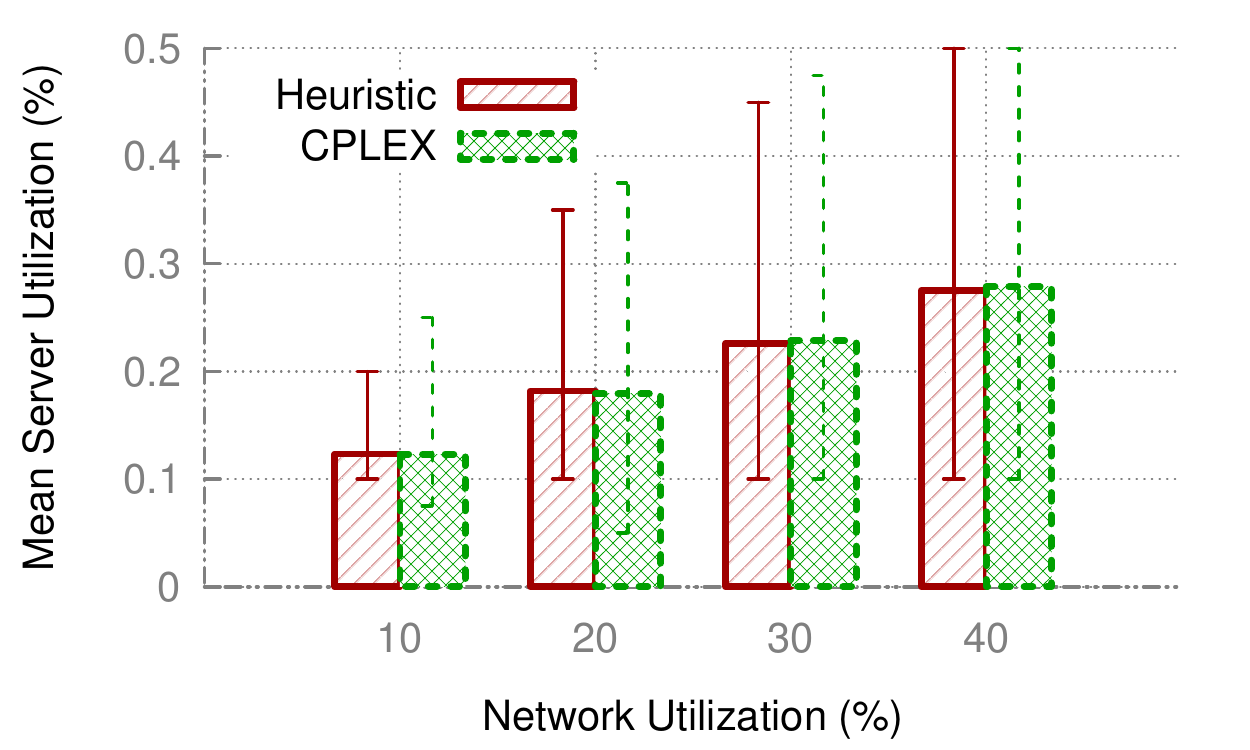}}
	\caption{Cost Ratio (Heuristic / CPLEX) with Varying Load}
	\label{fig:load_variation}
\end{figure}

\subsection{Effect of High Traffic Volume}\label{subsec:load_test}
Now, we show the impact of higher traffic volume on our solutions. We perform this experiment by increasing the original traffic by $10\%$ to $40\%$ (in increments of $10\%$) for the Internet2 topology (\fig{fig:load_variation}). We observed a linear relationship between OPEX and network utilization for both of our solutions. The cost also grows almost at the same rate for both CPLEX and heuristic as evident from~\fig{fig:cost_load_variation}. The heuristic is able to follow the optimal solution very closely. Although it might seem a bit unintuitive by looking at the ratio of the individual cost components, it occurs as the transit cost is two order-of-magnitude larger than the energy cost. 

The server utilization increases sub-linearly with increasing network load (\fig{fig:util_load_variation}). The number of used servers remains the same for different network loads, but more cores were used since more VNFs were deployed. The larger error bar for CPLEX indicates the deployment of more VNFs, which increases the energy cost. However, more VNFs eventually decreased the transit cost, which is the major contributor to OPEX in this case.


\section{Related Works}\label{sec:related_work}
The initial drive for NFV was from several telecommunication operators back in 2012~\cite{etsinfv}. The motivation behind NFV is to break the barrier of proprietary hardwares and have more flexibility in the network in terms of the placement of service points, introducing new services, vendor independence \etc\ To this date, research efforts have been made in different aspects of NFV. In this section, we discuss about state of the art NFV management and orchestration proposals (\sect{subsec:mannfv}) followed by some enabling technologies for NFV (\sect{subsec:entecnfv}).

\subsection{Management and Orchestration of Network Functions}\label{subsec:mannfv}
Some of the early works on managing Network Functions (NFs) in networks, propose to outsource them to a cloud service~\cite{sherry2012making, gibb2012outsourcing}. Such outsourcing is motivated in the literature by studying experiences of different network operators. \cite{sherry2012making, gibb2012outsourcing} show how the management complexities arising in today's enterprise networks can be mitigated by outsourcing.

A more management approach towards NFV is taken by projects like Stratos~\cite{gember2013stratos}, OpenNF~\cite{gember2014opennf}. Stratos proposes an architecture for orchestrating VNFs outsourced to a remote cloud by taking care of traffic engineering, horizontal scaling of VNFs \emph{etc.} On the other hand, OpenNF proposes a converged control plane for VNFs and network forwarding plane by extending the centralized SDN paradigm. Some recent research works~\cite{mehraghdam2014specifying,moensvnf} provide initial study on placing VNFs. However, none of the aforementioned research works address the issue of dynamically adjusting the placement of VNFs to balance between network operating cost and performance.

Some recent works on managing NFs focus on traffic engineering issues such as steering the traffic through some predefined sequence of NFs. This problem becomes more challenging when NFs along the sequence modify the packet headers, thus changing the traffic signature. Proposals like~\cite{fayazbakhsh2014enforcing} and~\cite{qazi2013simple} propose SDN based solutions to the traffic steering problem. They propose tagging based mechanisms to identify a traffic during its lifetime and also to keep track of the visited sequence of middleboxes. The global network view of SDN makes it easier to manage and assign tags to different traffics and to ensure different policy enforcement on NFs. 

\subsection{Enabling Technologies for NFV}\label{subsec:entecnfv}
NFV proposes to run VNFs on commodity hardware as virtual appliances. This flexibility raises the question of performance. In recent years, a number of research efforts have been targeted to achieve near line speed network I/O throughput with commodity servers~\cite{dpdk, rizzo2012netmap}. Apart from accelerating the packets along the network I/O stack, more recent works have proposed changes to virtualization technologies to support the deployment of modular software NFs on lightweight VMs~\cite{martins2014clickos}. Hundreds of these VMs can be instantiated on a single physical machine within miliseconds to run different VNFs. Substantial research efforts are also being put towards programming models and deployment architecture for VNFs as well. CoMb~\cite{sekar2012design} and xOMB~\cite{anderson2012xomb} propose an extensible and consolidated framework for incrementally developing scalable middleboxes. Both of these works leverage the idea of reusable network processing pipelines for middlebox composition.
\section{Conclusion}\label{sec:conclusion}
Virtualized network functions provide a flexible way to deploy, operate and orchestrate network services with much less capital and operational expenses. Software middleboxes (\eg\ ClickOS) are rapidly catching up with hardware middlebox performance. Network operators are already opting for NFV based solutions. We believe that our model for dynamic VNF orchestration will have significant impact on middlebox management in the near future. Our model can be used to determine the optimal number of VNFs and to place them at the optimal locations to optimize network operational cost and resource utilization. Our trace driven simulations on the Internet2 research network demonstrate that network OPEX can be reduced by a factor of $4$ over hardware middleboxes through proper VNF orchestration. 

In this paper, we presented two solutions to the VNF orchestration problem: CPLEX based optimal solution for small networks and a heuristic for larger networks. We found that the heuristic produces solutions that are within $1.3$ times of the optimal solution, yet the execution-time is about $65$ to $3500$ times faster than that of the CPLEX solution. We intend to extend this work in a number of ways. We plan to extend our model for supporting both hardware and software middleboxes in the same network. We want to explore the possibility of introducing failure-resilience by deploying backup VNFs that can take over the traffic processing tasks from failed VNFs. We also plan to enhance the physical network transformation process to further reduce the solution space and speed-up the running time of the optimal solution.
\bibliographystyle{abbrv}

\begin{thebibliography}{10}

\bibitem{serverpower}
Comparison of enterprise class power enclosure.
\newblock
  {http://www.dell.com/downloads/global/products/pedge/en/bladepower-\\studywhitepaper\_08112010\_final.pdf}.

\bibitem{sfcdraft}
https://datatracker.ietf.org/doc/draft-ietf-sfc-dc-use-cases/.

\bibitem{dpdk}
{Intel DPDK}.
\newblock http://dpdk.org/.

\bibitem{transitcost}
{Internet Transit Pricing}.
\newblock
  http://drpeering.net/white-papers/Internet-Transit-Pricing-Historical-And-Projected.php.

\bibitem{internet2data}
{Internet2 Research Network Topology and Traffic Matrix}.
\newblock http://www.cs.utexas.edu/\~{}yzhang/research/AbileneTM/.

\bibitem{vnfguide}
{pfSense Hardware Sizing Guide}.
\newblock {https://www.pfsense.org/hardware/\#sizing}.

\bibitem{anderson2012xomb}
J.~W. Anderson, R.~Braud, R.~Kapoor, G.~Porter, and A.~Vahdat.
\newblock {xOMB: Extensible open middleboxes with commodity servers}.
\newblock In {\em Proc. of ACM/IEEE ANCS '12}, pages 49--60.

\bibitem{benson2010network}
T.~Benson, A.~Akella, and D.~A. Maltz.
\newblock {Network traffic characteristics of data centers in the wild}.
\newblock In {\em Proc. of ACM IMC '10}, pages 267--280.

\bibitem{chiou2008transshipment}
C.-C. Chiou.
\newblock {Transshipment problems in supply chain systems: review and
  extensions}.
\newblock {\em Supply Chain}, pages 427--448, 2008.

\bibitem{Mosharaf10}
N.~M.~K. Chowdhury and R.~Boutaba.
\newblock A survey of network virtualization.
\newblock {\em Computer Networks}, 54(5):862 -- 876, 2010.

\bibitem{etsinfv}
ETSI.
\newblock {Network Functions Virtualisation -- Introductory White Paper}.
\newblock https://portal.etsi.org/NFV/NFV\_White\_Paper.pdf, 2012.

\bibitem{fayazbakhsh2014enforcing}
S.~K. Fayazbakhsh, L.~Chiang, V.~Sekar, M.~Yu, and J.~C. Mogul.
\newblock {Enforcing network-wide policies in the presence of dynamic middlebox
  actions using flowtags}.
\newblock In {\em Proc. of USENIX NSDI '14}.

\bibitem{forney1973viterbi}
G.~D. Forney~Jr.
\newblock {The Viterbi Algorithm}.
\newblock {\em Proc. of the IEEE}, 61(3):268--278, 1973.

\bibitem{gember2013stratos}
A.~Gember, A.~Krishnamurthy, S.~S. John, R.~Grandl, X.~Gao, A.~Anand,
  T.~Benson, V.~Sekar, and A.~Akella.
\newblock {Stratos: A Network-Aware Orchestration Layer for Virtual Middleboxes
  in Clouds}.
\newblock {\em arXiv preprint arXiv:1305.0209}, 2013.

\bibitem{gember2014opennf}
A.~Gember-Jacobson, R.~Viswanathan, C.~Prakash, R.~Grandl, J.~Khalid, S.~Das,
  and A.~Akella.
\newblock {OpenNF: enabling innovation in network function control}.
\newblock In {\em Proc. of ACM SIGCOMM '14}, pages 163--174.

\bibitem{gibb2012outsourcing}
G.~Gibb, H.~Zeng, and N.~McKeown.
\newblock {Outsourcing network functionality}.
\newblock In {\em Proc. of ACM HotSDN '12}, pages 73--78.

\bibitem{haeffner2014mobile}
W.~Haeffner, J.~Napper, M.~Stiemerling, D.~Lopez, and J.~Uttaro.
\newblock {Service Function Chaining Use Cases in Mobile Networks}, 2014.

\bibitem{179739}
J.~Hwang, K.~K. Ramakrishnan, and T.~Wood.
\newblock {NetVM: High Performance and Flexible Networking Using Virtualization
  on Commodity Platforms}.
\newblock In {\em Proc. of USENIX NSDI '14}, pages 445--458.

\bibitem{kou1977multidimensional}
L.~T. Kou and G.~Markowsky.
\newblock {Multidimensional bin packing algorithms}.
\newblock {\em IBM Journal of Research and development}, 21(5):443--448, 1977.

\bibitem{kreutz2014software}
D.~Kreutz, F.~Ramos, P.~Verissimo, C.~E. Rothenberg, S.~Azodolmolky, and
  S.~Uhlig.
\newblock {Software-Defined Networking: A Comprehensive Survey}.
\newblock {\em arXiv preprint arXiv:1406.0440}, 2014.

\bibitem{liu2014service}
W.~Liu, H.~Li, O.~Huang, M.~Boucadair, N.~Leymann, Q.~Fu, Q.~Sun, C.~Pham,
  C.~Huang, J.~Zhu, and P.~He.
\newblock {Service Function Chaining Problem Statement}.
\newblock {\em draft-liu-sfc-use-cases-08 (work in progress)}, 2014.

\bibitem{martins2014clickos}
J.~Martins, M.~Ahmed, C.~Raiciu, V.~Olteanu, M.~Honda, R.~Bifulco, and
  F.~Huici.
\newblock {ClickOS and the art of network function virtualization}.
\newblock In {\em Proc. of USENIX NSDI '14}, pages 459--473.

\bibitem{mehraghdam2014specifying}
S.~Mehraghdam, M.~Keller, and H.~Karl.
\newblock {Specifying and Placing Chains of Virtual Network Functions}.
\newblock {\em arXiv preprint arXiv:1406.1058}, 2014.

\bibitem{moensvnf}
H.~Moens and F.~De~Turck.
\newblock {VNF-P: A Model for Efficient Placement of Virtualized Network
  Functions}.
\newblock In {\em Proc. of ManSDN/NFV '14}.

\bibitem{nucci2005problem}
A.~Nucci, A.~Sridharan, and N.~Taft.
\newblock The problem of synthetically generating ip traffic matrices: initial
  recommendations.
\newblock {\em ACM CCR}, 35(3):19--32, 2005.

\bibitem{Pirkul:1998:MMC:300685.300697}
H.~Pirkul and V.~Jayaraman.
\newblock {A Multi-commodity, Multi-plant, Capacitated Facility Location
  Problem: Formulation and Efficient Heuristic Solution}.
\newblock {\em Comput. Oper. Res.}, 25(10):869--878, Oct. 1998.

\bibitem{qazi2013simple}
Z.~A. Qazi, C.-C. Tu, L.~Chiang, R.~Miao, V.~Sekar, and M.~Yu.
\newblock {SIMPLE-fying middlebox policy enforcement using SDN}.
\newblock In {\em Proc. of ACM SIGCOMM '13}, pages 27--38.

\bibitem{quinn2014service}
P.~Quinn and T.~Nadeau.
\newblock {Service Function Chaining Problem Statement}.
\newblock {\em draft-quinn-sfc-problem-statement-10 (work in progress)}, 2014.

\bibitem{rizzo2012netmap}
L.~Rizzo.
\newblock {netmap: A Novel Framework for Fast Packet I/O.}
\newblock In {\em Proc. of USENIX ATC '12}, pages 101--112.

\bibitem{fnss}
L.~Saino, C.~Cocora, and G.~Pavlou.
\newblock {A Toolchain for Simplifying Network Simulation Setup}.
\newblock In {\em Proc. of SIMUTOOLS '13}.

\bibitem{sekar2012design}
V.~Sekar, N.~Egi, S.~Ratnasamy, M.~K. Reiter, and G.~Shi.
\newblock {Design and Implementation of a Consolidated Middlebox Architecture.}
\newblock In {\em Proc. of USENIX NSDI '12}, pages 323--336.

\bibitem{sherry2012making}
J.~Sherry, S.~Hasan, C.~Scott, A.~Krishnamurthy, S.~Ratnasamy, and V.~Sekar.
\newblock {Making middleboxes someone else's problem: network processing as a
  cloud service}.
\newblock {\em ACM CCR}, 42(4):13--24, 2012.

\bibitem{Sherry:EECS-2012-24}
J.~Sherry and S.~Ratnasamy.
\newblock {A Survey of Enterprise Middlebox Deployments}.
\newblock Technical Report UCB/EECS-2012-24, EECS Department, University of
  California, Berkeley, Feb 2012.

\bibitem{spring2002measuring}
N.~Spring, R.~Mahajan, and D.~Wetherall.
\newblock {Measuring ISP topologies with Rocketfuel}.
\newblock 32(4):133--145, 2002.

\bibitem{surendra2014dc}
S.~Surendra, M.~Tufail, S.~Majee, C.~Captari, and S.~Homma.
\newblock {Service Function Chaining Use Cases in Mobile Networks}, 2014.

\end{thebibliography}
{

}
\appendices
\begin{figure*} [!t]
	\centering
	\subfigure[Example Network Topolog]{\includegraphics[width=0.275\textwidth]{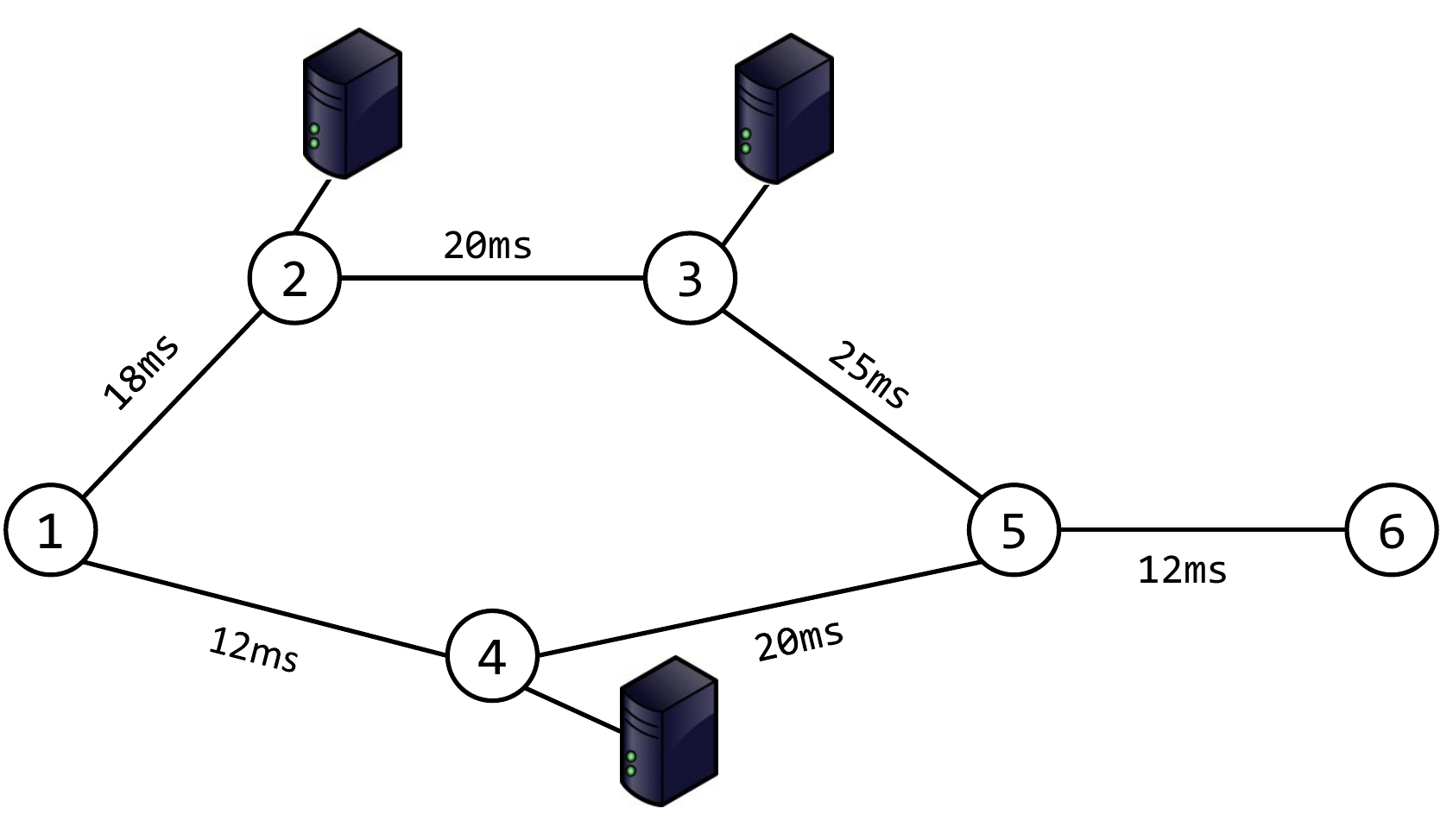}\label{fig:example_topology}}
	\subfigure[Multi-Stage Graph]{\includegraphics[width=0.275\textwidth]{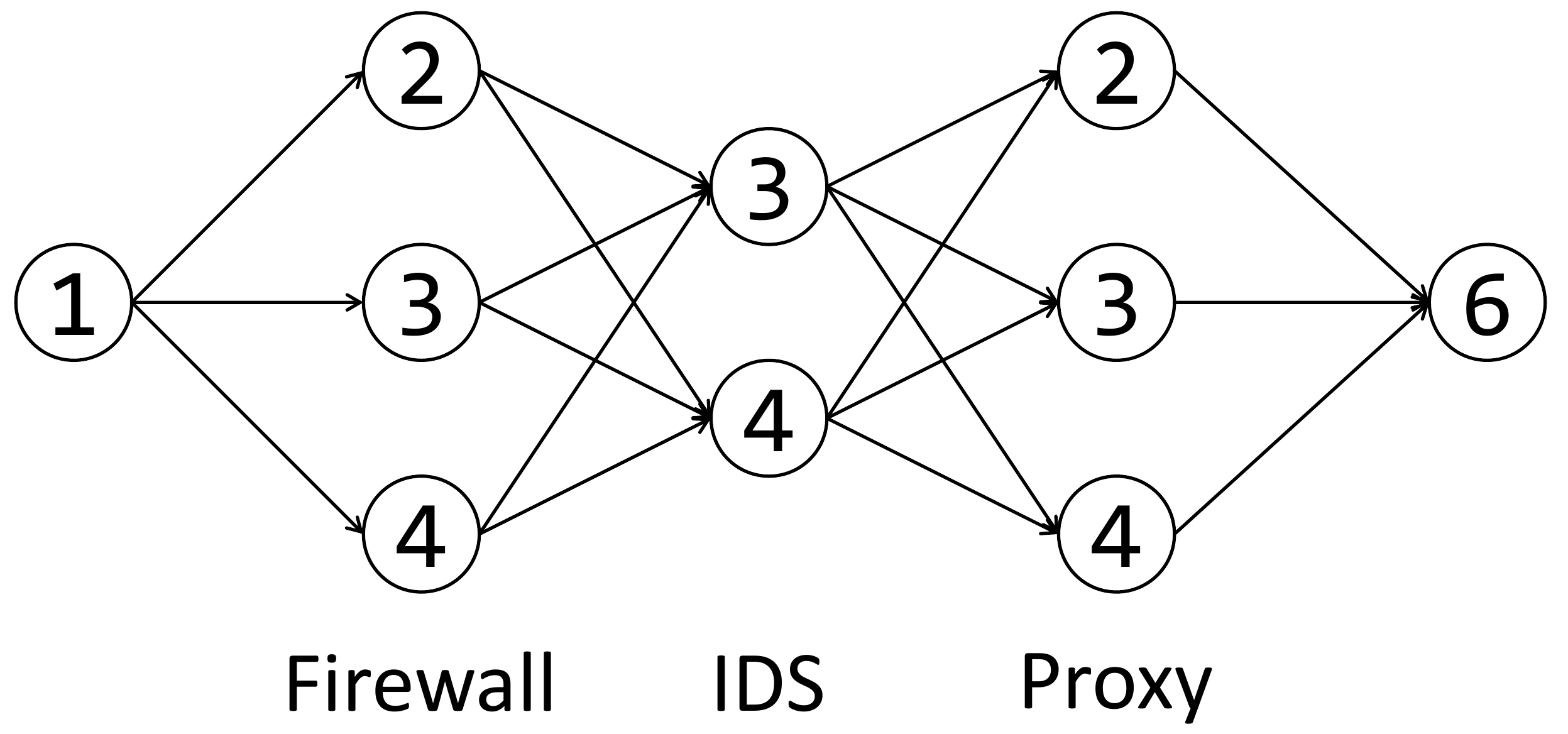}\label{fig:trellis_diagram}}
	\subfigure[A Single Stage]{\includegraphics[width=0.275\textwidth]{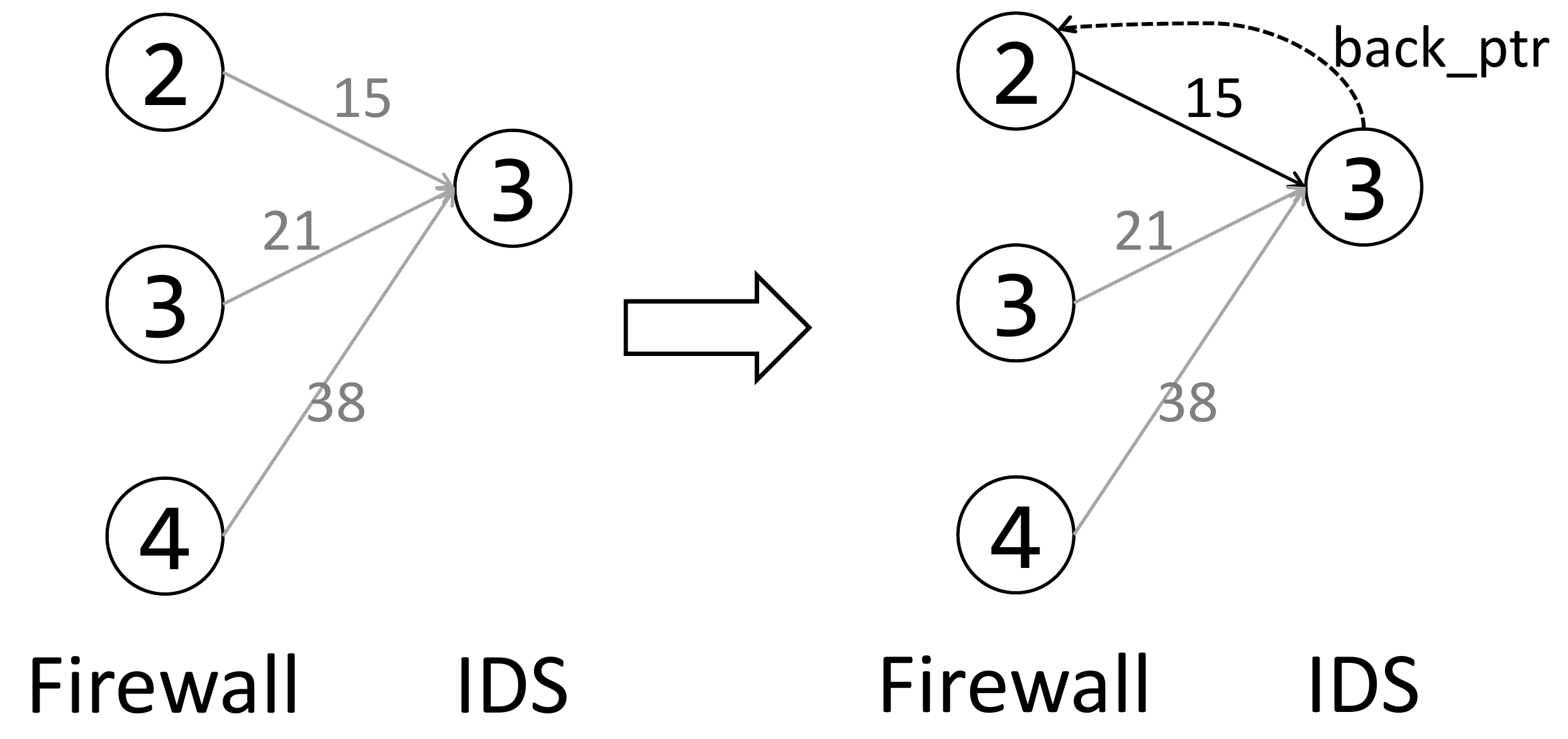}\label{fig:a_stage}}
	\caption{Modeling with Multi-Stage Graph}
	\label{fig:heuristic_in_action}
\end{figure*}

\section{Glossary of Symbols}

\begin{table}[!h]
  \label{tab:symbols}
  \centering
  \begin{tabular}{|l|l|}
  \hline
  \multicolumn{2}{|c|}{\textbf{Physical Network}}\\
  \hline
  $\bar{G}(\bar{S}, \bar{L})$ & Physical network $\bar{G}$  with switches $\bar{S}$ and links $\bar{L}$\\
  \hline
  $\bar{N}$ & Set of servers\\
  \hline
  $\bar{h}_{\bar{n}\bar{s}} \in \{0, 1\}$ & If server $\bar{n} \in \bar{N}$ is attached to switch $\bar{s} \in \bar{S}$\\
  \hline
  $R$ & Set of resources offered by servers\\
  \hline
  $c_{\bar{n}}^r \in \mathbb{R}^+$ & Resource capacity of server $\bar{n},\ \forall r \in R$\\
  \hline
  $\beta_{\bar{u}\bar{v}}, \delta_{\bar{u}\bar{v}} \in \mathbb{R}^+$ & Bandwidth, propagation delay of link $(\bar{u}, \bar{v}) \in \bar{L}$\\
  \hline
  $\eta(\bar{u})$ & Neighbors of switch $\bar{u}$\\
  \hline
  \multicolumn{2}{|c|}{\textbf{Virtualized Network Functions (VNFs)}}\\
  \hline
  $P$  & Set of possible VNF types\\
  \hline
  $\mathcal{D}_p^+, \kappa_p^r, c_p, \delta_p$ & Deployment cost, resource requirement, processing\\
   & capacity and processing delay of VNF type $p \in P$\\
  \hline
  $d_{\bar{n}p} \in \{0, 1\}$  & If VNF type $p$ can be provisioned on server $\bar{n}$\\
  \hline
  \multicolumn{2}{|c|}{\textbf{Traffic}}\\
  \hline
  $\bar{u}^t, \bar{v}^t, \Psi^t$ & Ingress, egress and VNF sequence for traffic $t$\\
  \hline
  $\beta^t, \delta^t, \omega^t$ & Bandwidth, expected delay, SLA penalty for $t$\\
  \hline
  $N^t$ & $\{\bar{u}^t, \bar{v}^t, \Psi^t\}$\\
  \hline
  $L^t$ & $\{(\bar{u}^t, \Psi^t_1), \ldots, (\Psi^t_{|\Psi^t|-1}, \Psi^t_{|\Psi^t|}), (\Psi^t_{|\Psi^t|}, \bar{v}^t)\}$\\
  \hline
  $\eta^t(n)$ & Neighbors of $n \in N^t$\\
  \hline
  $g_{np}^t \in \{0, 1\}$ & If node $n \in N^t$ is of type $p \in P$\\
  \hline
  $\mathcal{M}$ & Set of pseudo-VNFs\\
  \hline
  $\zeta(m)$ & $\zeta(m) = \bar{n} \text{ if VNF } m \in \mathcal{M} \text{ is attached to server } \bar{n}$\\
  \hline
  $\Omega(\bar{n})$ & $\{ m\ |\ \zeta(m) = \bar{n} \},\ m \in \mathcal{M}, \bar{n} \in \bar{N}$\\
  \hline
  $q_{mp} \in \{0, 1\}$ & If VNF $m \in \mathcal{M}$ is of type $p \in P$\\
  \hline
  \multicolumn{2}{|c|}{\textbf{Decision Variables}}\\
  \hline
  *$x_{nm}^t \in \{0, 1\}$ & If node $n \in N^t$ is provisioned on $m \in \mathcal{M}$\\ 
  \hline
  *$w_{\bar{u}\bar{v}}^{tn_{1}n_{2}}$ & $\text{If } (n1, n2) \in L^t \text{ uses physical link } (\bar{u}, \bar{v}) \in \bar{L}$\\
  \hline
  \multicolumn{2}{|c|}{\textbf{Derived Variables}}\\
  \hline
  *$y_m \in \{0, 1\}$ & If VNF $m \in \mathcal{M}$ is active\\
  \hline
  $z_{n\bar{s}}^t \in \{0, 1\}$ & If node $n \in N^t$ is attached to switch $\bar{s}$\\
  \hline
  \multicolumn{2}{|l|}{*$\hat{x}_{nm}^t, \hat{w}_{\bar{u}\bar{v}}^{tn_{1}n_{2}}, \hat{y}_m$ denote value from the previous iteration}\\
  \hline
  \end{tabular}
\end{table}

\section{Heuristic Algorithm}\label{appsec:algorithm}
Algorithm~\ref{alg:place-middlebox} gives the pseudcode of the heuristic solution. The procedure $ProvisionTraffic$ takes as input a traffic request $t$ and the topology graph $\bar{G}$ annotated with the resource capacities at each switch. We keep two tables, $cost$ and $\pi$, to keep track of the cost and the sequence of middlebox placements, respectively. $cost_{i,j}$ represents the cost of deploying the $j$-th middlebox in the middlebox sequence $\Psi^t$ to a server attached with switch $i$. The cost computation procedure is the same as described in~\sect{subsec:viterbi}. We use a number of helper procedures for the ease of implementation. The first helper procedure, $IsResourceAvailable$ checks if a middlebox $mbox$ for a traffic request $t$ can be placed at a switch $i$, satisfying the minimum bandwidth and resource requirements. The second helper, $GetCost$, computes the cost of placing middlebox $mbox$ for a traffic request $t$ at a server attached to switch $j$. The previous node $k$ that yields the minimum cost for the current node in consideration $j$, is tracked by the entry $\pi_{k,j}$. Finally, we backtrace using entries in $\pi$ to obtain the desired middlebox sequence. 

\textbf{Running Time: } Let the number of switches and the maximum length of a middlebox sequence be $n$ and $m$, respectively. Algorithm~\ref{alg:place-middlebox} performs $\Theta(nm)$ computations at the beginning to initialize the cost matrix. Then for each element in the traffic sequence, the algorithm takes all possible pairs of nodes $u$, $v$ and computes the cost of deploying a middlebox at the server attached to switch $v$ given that the previous middlebox in the sequence was deployed at a server connected to switch $u$. Therefore, there is a total of $\Theta(n^2m)$ operations involved. With some pre-computation steps the costs can be calculated and resource availability can be queried in $O(1)$ time. Therefore, Algorithm~\ref{alg:place-middlebox} runs in $\Theta(n^2m)$.

\begin{figure}
\centering
  \includegraphics[width=0.43\textwidth]{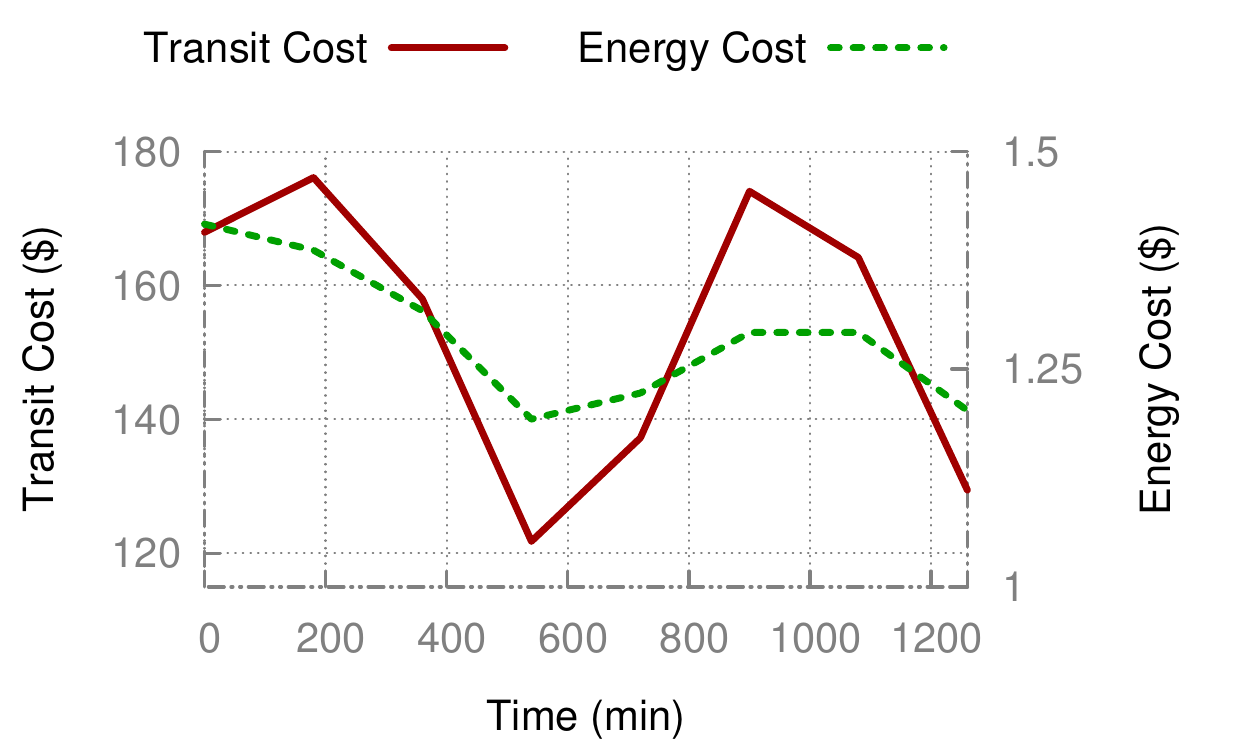}
  \caption{OPEX Components for AS-3967}
  \label{fig:3967_opex}
\end{figure}

\begin{figure*}
\centering
	\subfigure[Mean Server Utilization ]{\label{fig:util_ts_3967}\includegraphics[width=0.245\textwidth]{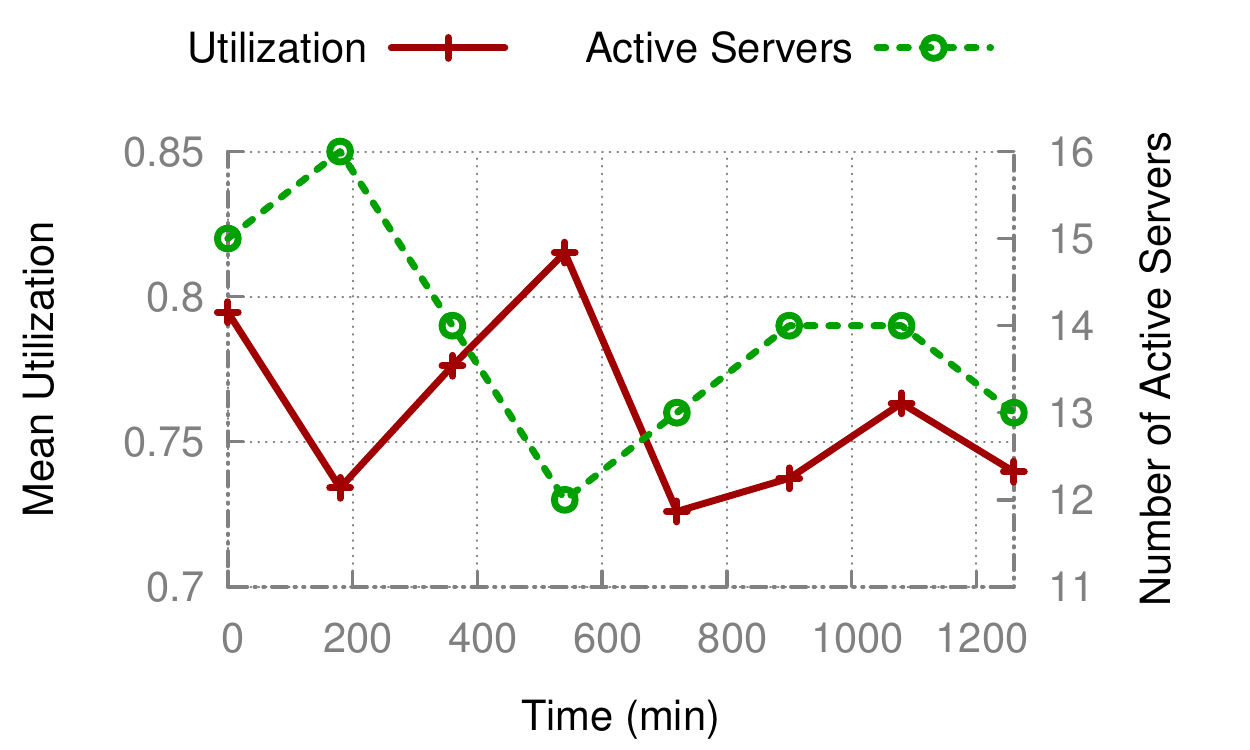}}
	\subfigure[Per Server Utilization ]{\label{fig:util_ps_3967}\includegraphics[width=0.245\textwidth]{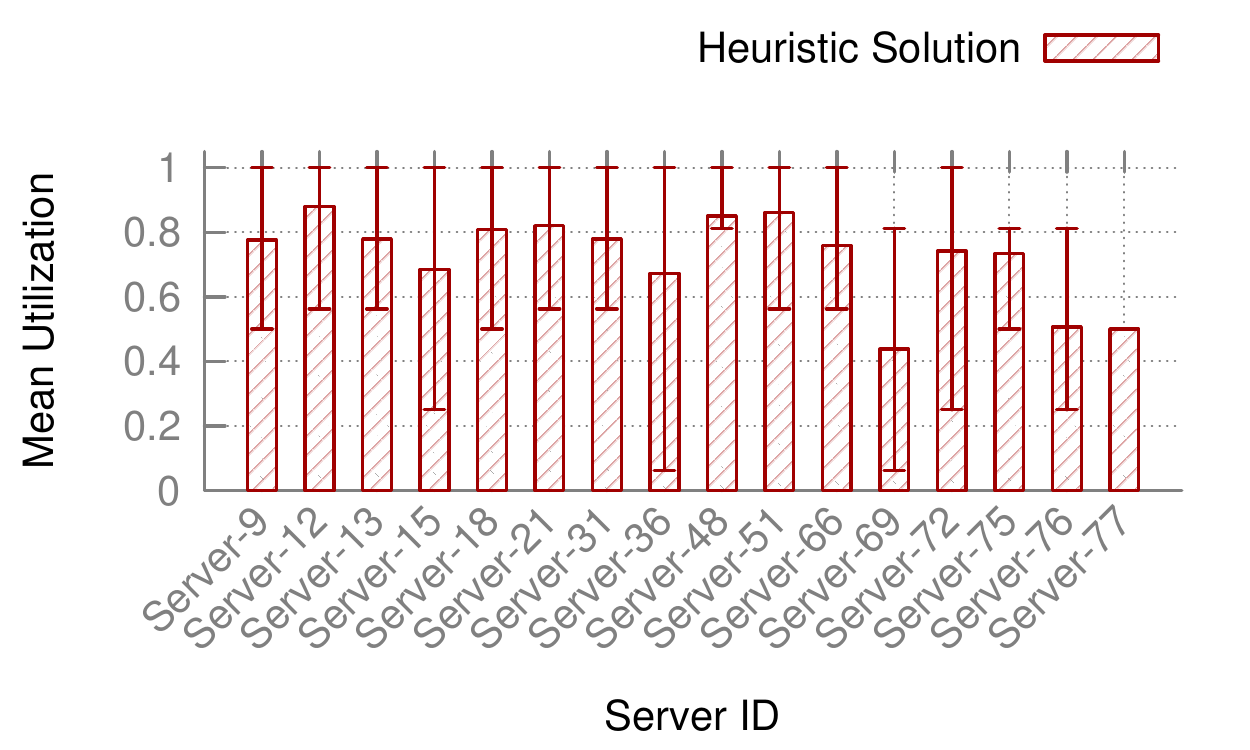}}
	\subfigure[Distance to middlebox]{\label{fig:ingress_3967}\includegraphics[width=0.245\textwidth]{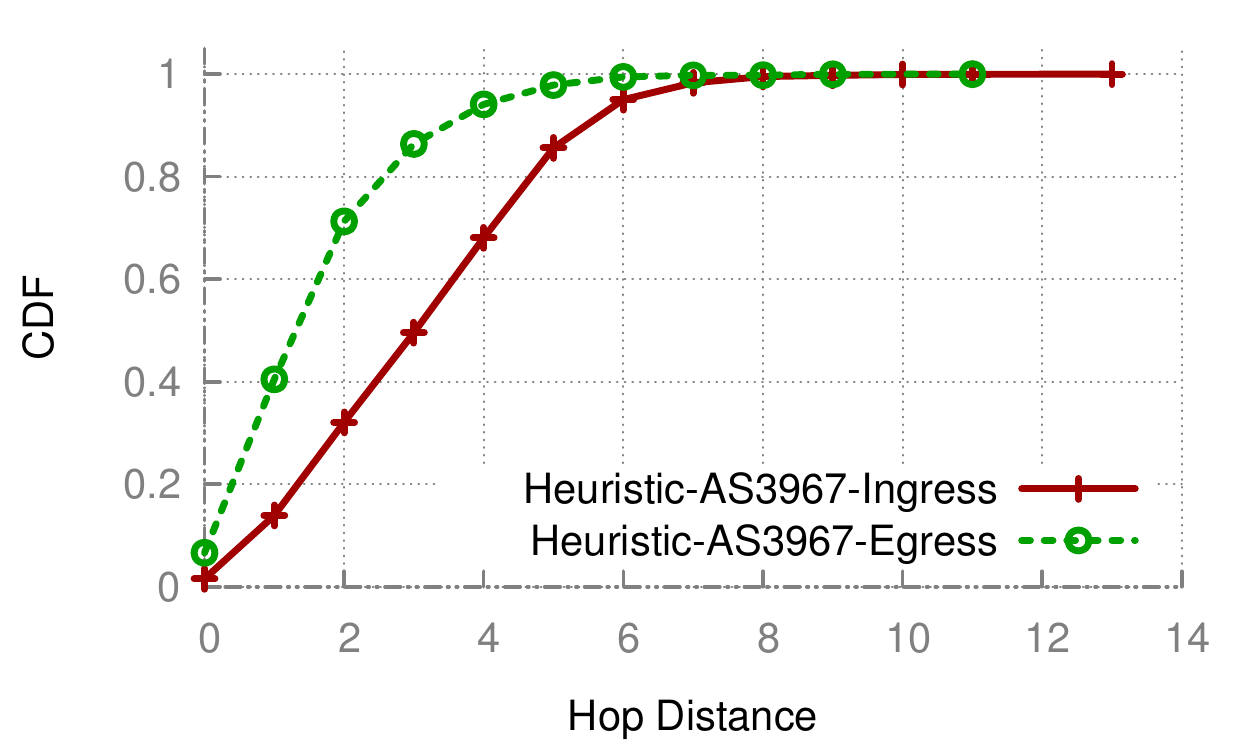}}
	\subfigure[Path stretch]{\label{fig:stretch_3967}\includegraphics[width=0.245\textwidth]{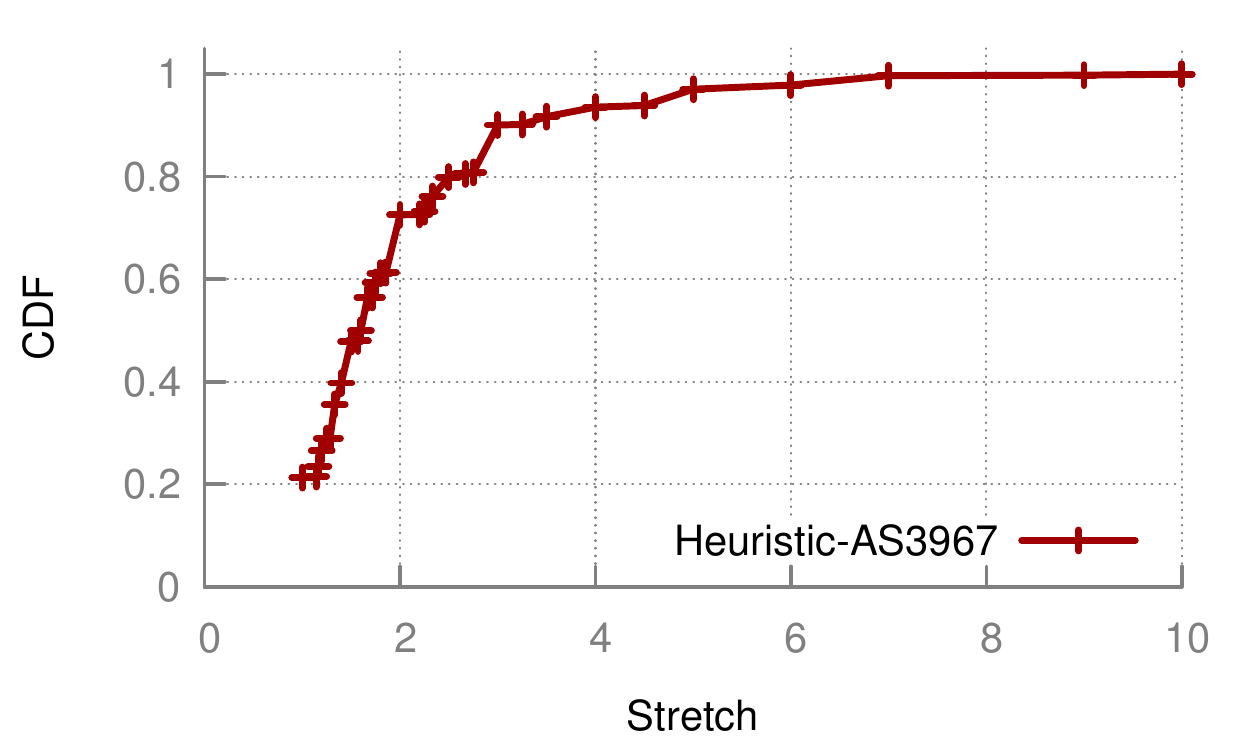}}
	\caption{Results for Rocketfuel Topology (AS-3967)}
	\label{fig:as-3967}
\end{figure*}

\begin{figure*}
\centering
	\subfigure[Internet2]{\label{fig:internet2}\includegraphics[width=0.3\textwidth]{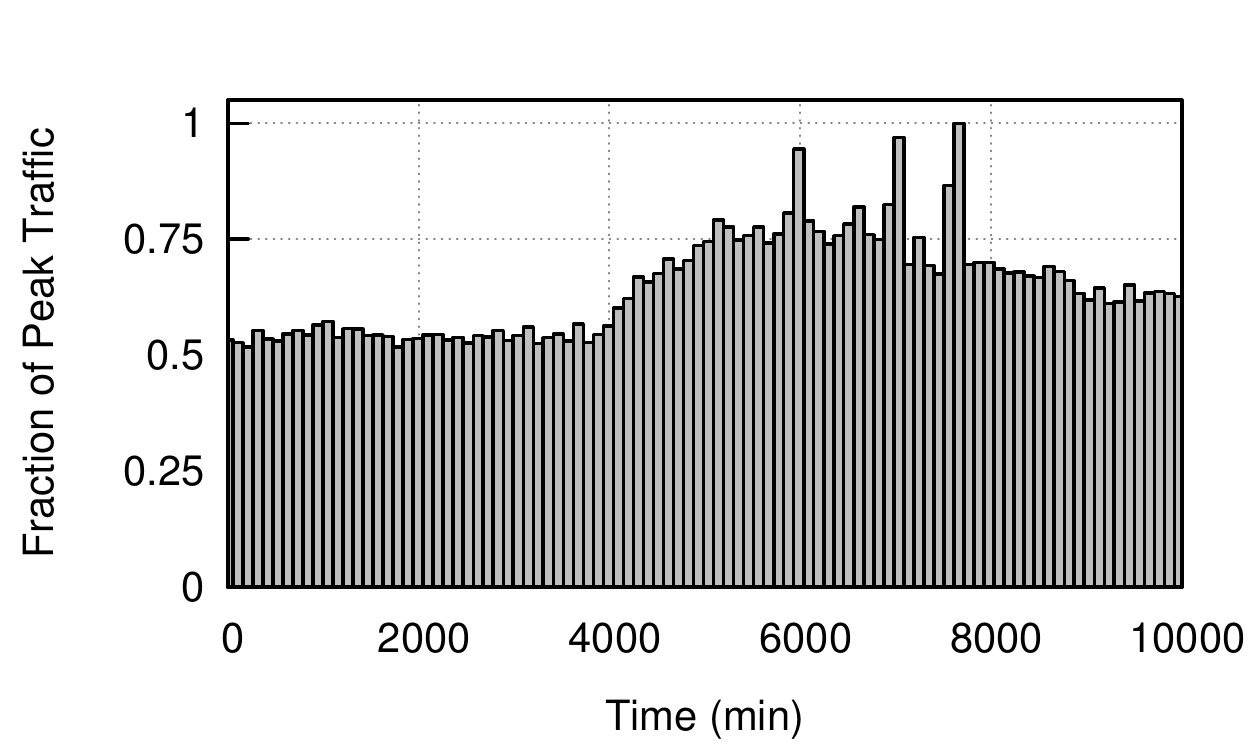}}
	\subfigure[Data Center]{\label{fig:dc}\includegraphics[width=0.3\textwidth]{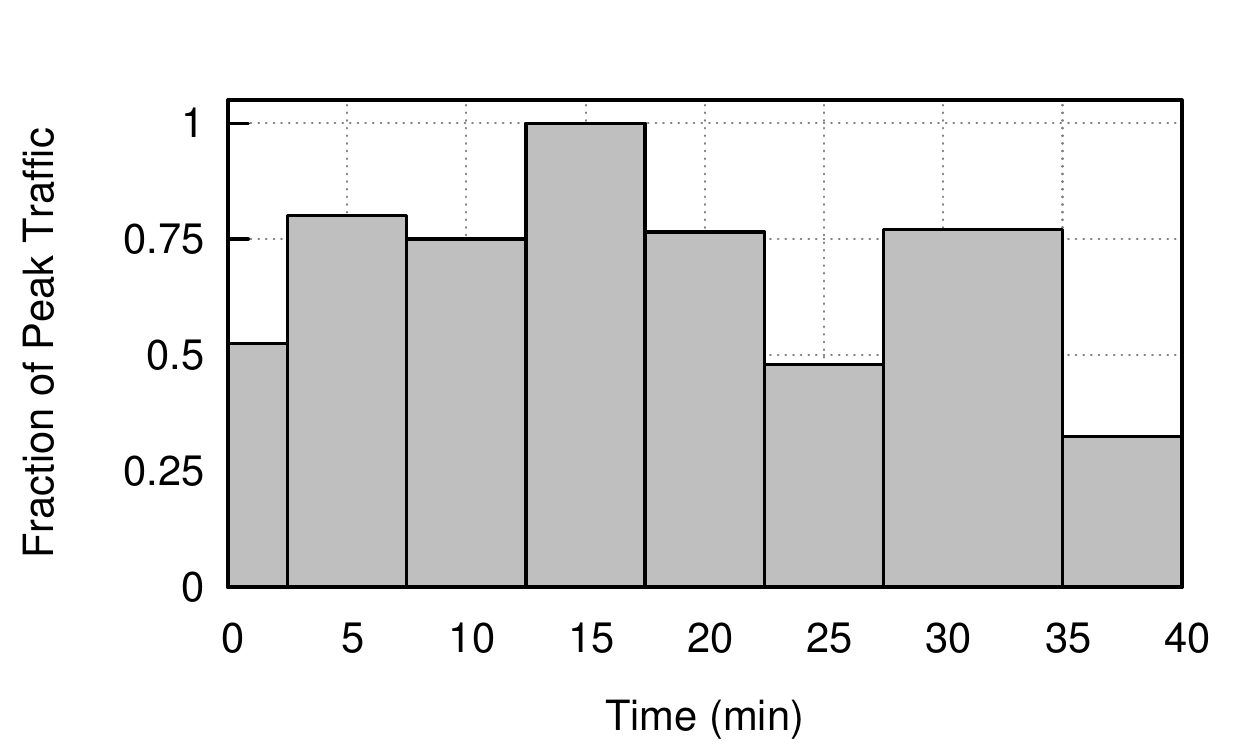}}
	\subfigure[Rocketfuel]{\label{fig:rf_traffic}\includegraphics[width=0.3\textwidth]{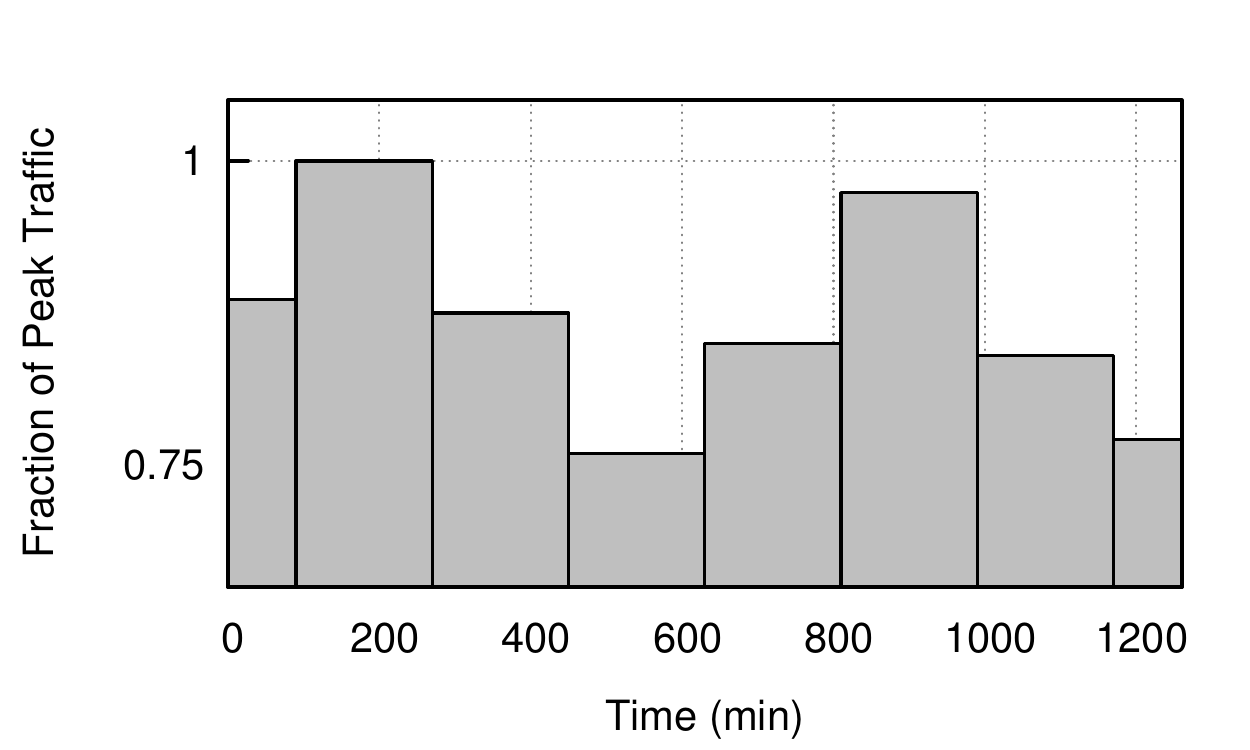}}
	\caption{Traffic Distribution over Time for Different Scenarios}
\end{figure*}

\section{Heuristic in Action}\label{appsec:example}
\fig{fig:example_topology} shows an example network topology with six switches, where the servers are connected to switch 2, 3 and 4. Now, we are required to find the path for a traffic which is going from switch 1 to 6 and must pass through a firewall, then an IDS and finally through a proxy.

Now, we generate a multi-stage graph as shown in~\fig{fig:trellis_diagram}. Here, we are assuming that the firewall and proxy can be deployed on any server, but the IDS can only be deployed on servers connected to switch 3 and 4. Each node in the multi-stage graph represents a decision about where to place a VNF. For example, if we select node 4 in the stage labeled ``IDS'', it means that a VNF corresponding to an IDS will be deployed on the server connected to switch 4. As explained earlier, there is a cost associated with each node selection. 

Now, we traverse this graph starting at node 1. The first stage is trivial, we just compute the cost of deploying and running (energy cost) a firewall at node 2, 3 and 4 and add the cost of routing traffic from node 1 to each node. There is no additional computation as there is just one incoming link for each node. However, the operations for the subsequent stages involve comparing the cost of reaching a particular node from different nodes. For example, node 3 in stage ``IDS'' can be reached from three different nodes. The operation performed in this is stage is explained in~\fig{fig:a_stage}.

We need to compute the cost of transition from nodes 2, 3 and 4 to node 3. These costs are shown on the left side of~\fig{fig:a_stage}. Now, if we select the link between node 4 and node 3 then the Firewall will be deployed on node 4 and the IDS will be deployed on node 3 and cost of deploying the IDS will be 38. However, we have links with lower costs than this one and at each stage we select the incoming link with the minimal cost. So, here we will select the link between node 2 and 3 as it has the lowest cost of 15. We will also save a pointer ($\texttt{back\_ptr}$) to mark the node that was selected. We continue in this manner until we reach the destination node (node 6 in this example), then we follow the $\texttt{back\_ptr}$s to re-construct the solution. 

\section{Results for Rocketfuel Topology}
The results for the AS-3967 topology is shown in~\fig{fig:3967_opex} and~\fig{fig:as-3967}. The traffic for this topology is show in~\fig{fig:rf_traffic}. As mentioned earlier, this traffic was generated using the FNSS tool~\cite{fnss} and it exhibits time-of-day effect. We cannot provide a comparison with the optimal solution as the CPLEX program was not able to solve the problem for this topology. It failed to fit the optimization model in its memory even though the physical machine had 1TB of memory. The program crashes after the total memory usage reaches around 300 GB. We observed similar behavior when experimenting with high traffic volumes. CPLEX was not able to solve the problem for the Internet2 topology when traffic was increased to utilize the network by more than $40\%$. We tuned different parameters (\eg\ solving the dual problem, storing branch and bound tree data on disk, reducing the number of threads, \etc) of the CPLEX solver according to the guidelines provided by IBM\footnote{http://www-01.ibm.com/support/docview.wss?uid=swg21399933}, but could not solve the problem. We plan to investigate this issue further in the future. However, the heuristic solution was able to solve the same problem in less than 3 seconds.

The transit and energy cost for the AS-3967 topology is reported in~\fig{fig:3967_opex}. The transit cost is two order-of-magnitude higher than the energy cost, which is expected for a larger network with large amount of traffic. From~\fig{fig:rf_traffic} and~\fig{fig:3967_opex}, we can see that our dynamic VNF orchestration approach adapts nicely with the changing traffic conditions. It can dynamically scale-up or scale-down the number of active VNFs (demonstrated by the rise and fall of the energy cost). It can also adapt the location of the VNFs according to the variation in the traffic volume. 

The results for system resource utilization and topological properties for middlebox locations are shown in~\fig{fig:as-3967}. From~\fig{fig:util_ts_3967} we can see that the mean utilization and number of active servers vary with fluctuation in traffic volume. The mean utilization of the servers is around $80\%$, but there is a small number of servers that are underutilized (\fig{fig:util_ps_3967}). The CDF of percentage of middleboxes deployed within $k$-hop distance from the ingress switch is reported in~\fig{fig:ingress_3967}. More than $90\%$ middleboxes are deployed within $5$ hops, which is quite reasonable for a network with $79$ switches and $147$ links. Similar results are obtained for the egress case as shown in the same figure. Finally, the path stretch is shown in~\fig{fig:stretch_3967}. We can observe that $20\%$ traffic passes through the shortest path even after going though the VNF sequence. So, in $20\%$ of the cases VNFs are provisioned on the shortest path between the ingress and egress switches that the traffic is passing through. 

\end{document}